\UseRawInputEncoding
\documentclass[review,authoryear]{elsarticle}

\usepackage{etoolbox}
\makeatletter
\patchcmd{\ps@pprintTitle}% <cmd>
  {Elsevier}% <search>
  {journal (under review)}% <replace>
  {}{}% <succes><failure>
\makeatother

\usepackage[english]{babel}
\usepackage{graphicx}
\usepackage{amsmath}
\usepackage{amssymb}
\usepackage{esvect}
\usepackage{mathtools}
\usepackage{algorithm} 
\usepackage{algpseudocode} 

\usepackage{hyperref}
\usepackage{caption}
\usepackage{float}
\usepackage{array}
\usepackage{bm}
\usepackage{optidef}
\usepackage{color}
\usepackage{amsthm}	% NEEDED
\usepackage{comment}
\DeclareMathOperator{\sign}{sign}

\usepackage{natbib}
\usepackage{geometry}
\usepackage{fleqn}

\usepackage{multirow}

\algdef{SE}[SUBALG]{Indent}{EndIndent}{}{\algorithmicend\ }%
\algtext*{Indent}
\algtext*{EndIndent}

\title{Learning port maneuvers from data for automatic guidance \\ of Unmanned Surface Vehicles  \tnoteref{t1}}
\tnotetext[t1]{This work has been supported by the emerging research group Multi-robot And Control Systems (\href{https://prisma.us.es/colectivo/grupo/TEP-995}{MACS}) under the framework VII PPIT-US 2024, and by the National Program for Doctoral Formation (Minciencias-Colombia, 885-2020).}

\author[a,b]{Yeyson A. Becerra-Mora\corref{cor1}}
\ead{yeyson_becerra@cun.edu.co}
\author[a]{Jos\'e \'Angel Acosta}
\ead{jaar@us.es}
\author[a]{\'Angel Rodr\'iguez Casta\~no}
\ead{castano@us.es}

\cortext[cor1]{Corresponding author}
\affiliation[a]{organization={Dept. Ingenieria de Sistemas y Automatica, Universidad de Sevilla},
city={Sevilla},
postcode={41092},
country={Spain}}

\affiliation[b]{organization={Dept. of Electronic Engineering, CUN},
city={Bogota},
postcode={111711},
country={Colombia}}

\begin{document}

% make the title area
%\maketitle
%\thispagestyle{empty}
%\pagestyle{empty}

% As a general rule, do not put math, special symbols or citations
% in the abstract or keywords.
\begin{abstract}
At shipping ports, some repetitive maneuvering tasks such as entering/leaving port, transporting goods inside it or just making surveillance activities, can be efficiently and quickly carried out by a domestic pilot according to his experience. This know-how can be seized by Unmanned Surface Vehicles (USV) in order to autonomously replicate the same tasks. However, the inherent nonlinearity of ship trajectories and environmental perturbations as wind or marine currents make it difficult to learn a model and its respective control. We therefore present a data-driven learning and control methodology for USV, which is based on Gaussian Mixture Model, Gaussian Mixture Regression and the Sontag's universal formula. Our approach is capable to learn the nonlinear dynamics as well as guarantee the convergence toward the target with a robust controller. Real data have been collected through experiments with a vessel at the port of Ceuta. The complex trajectories followed by an expert have been learned including the robust controller. The effect of the controller over noise/perturbations are presented, a measure of error is used to compare estimates and real data trajectories, and finally, an analysis of computational complexity is performed.
\end{abstract}

\maketitle

\section{Introduction}
Unmanned Surface Vehicle (USV) have gained popularity across the world in recent years due to their efficiency and safety to carry out its operations; furthermore, financial losses as well as maritime accidents caused by human mistakes have been minimized. Applications such as people transportation or payload carry, surveillance and rescue, target recognition, among others can be performed by USV; nevertheless, most of the time humans are in charge of developing these kinds of tasks, gaining a large amount of experience from it. AI techniques can leverage this human experience to teach such tasks to the machines. 

USV trajectories are inherently nonlinear, hence it is needed to seek methods or strategies to deal with such dynamics. Methods like path-planning, reinforcement learning, Gaussian Process, among others have been proposed to control autonomous vessels. One of the most popular methods employed for USV is path-planning \cite{Wang20}; \cite{GONZALEZGARCIA}; \cite{XIANGENBAI}; \cite{BRUAL}; \cite{NINGWANG}; which allows to follow a trajectory through waypoints and sensors to identify the system state in real-time, and to avoid obstacles. Likewise, reinforcement learning through policies/rewards and waypoints is another option for these systems to learn trajectories \cite{DERAJ}, to perform collision avoidance \cite{PENGWANG}; \cite{NINGWANG}; or even to learn the swing process manipulation of a cutter suction dredger through human demonstrations \cite{WEI}. The USV generates complex dynamics in its trajectories that are difficult to learn, thus Gaussian Process (GP) methods turns an option to learn them as presented in \cite{OUYANG}; \cite{MENG}; \cite{YIFANXUE}.

Learning from Demonstrations (LfD) \cite{Calinon09book} is an employed method to teach tasks that are non-ease to program into an autonomous system; furthermore, a human expert in a specific task is capable to transmit his knowledge to a robot without writing a single line of code for it. This method is used to teach skills to manipulators \cite{VOGT}; biped robots \cite{FARCHY}; unmanned ground vehicles (UGV) \cite{LI2017}; unmanned aerial vehicles (UAV) \cite{LOQUERCIO}; and autonomous underwater vehicle (AUV) \cite{BIRK}. One of the main advantages about this method over path-planning or even reinforcement learning is having no necessity to discretize the trajectories as waypoints, neither an environment map once the dynamic is learned.

Most of the studies related to trajectory following with USV are based on path-planning and reinforcement learning; whilst methods based on LfD or Imitation Learning (IL) have reported scarce studies \cite{CHAYSRI}; \cite{ZHANG2023}. Sometimes, learning a USV trajectory is not enough to cope environment features like wind and oceanic currents, hence control techniques are essential to ensure USV stability. Combining learning methods and nonlinear control techniques yields more robust data-driven controllers for autonomous vessels like the one presented in \cite{PEILONG}; which is based on GP and Nonlinear Model Predictive Control (NMPC). 

A data-driven learning and control approach for USVs is presented in this research. Its main contributions are: 1) learning of position and heading; 2) great robustness, guaranteeing both fidelity and convergence toward the target despite disturbances in the marine environment; and 3) low computational cost, enabling the implementation of corrections during navigation. The trajectory demonstrations are performed by a human expert navigating at the port of Ceuta (Spain) and learned by the system from the real data collected through Gaussian Mixture Model (GMM). Then a Gaussian Mixture Regression is used to estimate real trajectories, and finally a control law based on Sontag's formula is proposed to reduce data measurement noise, perturbations from ocean currents and guarantee convergence to the target.

The paper is organized as follows. Section 2 presents the realm of port docking and employed USV. Section 3 introduces the core of the learning and control including: the fundamentals of GMM/GMR, data-driven control and the optimization problem. Section 4 describes the experimental validation of the proposed method. Finally, section 6 summarizes the main results and future research.

\noindent{\bf Notation.} $|\cdot|$ stands for vector norm. For a matrix $W$, $W\succ 0$ stands for symmetric and positive definite. For a scalar function $V(x)$, $x\in \mathbb{R}^{d}$, the gradient with respect to vector $x$ is denoted as $\nabla_{x} V(x)$. All vectors including the gradient are defined as columns.

\section{The realm of autonomous port docking}

\begin{figure}[htbp]
    \centering 
    \includegraphics[width=0.62\textwidth]{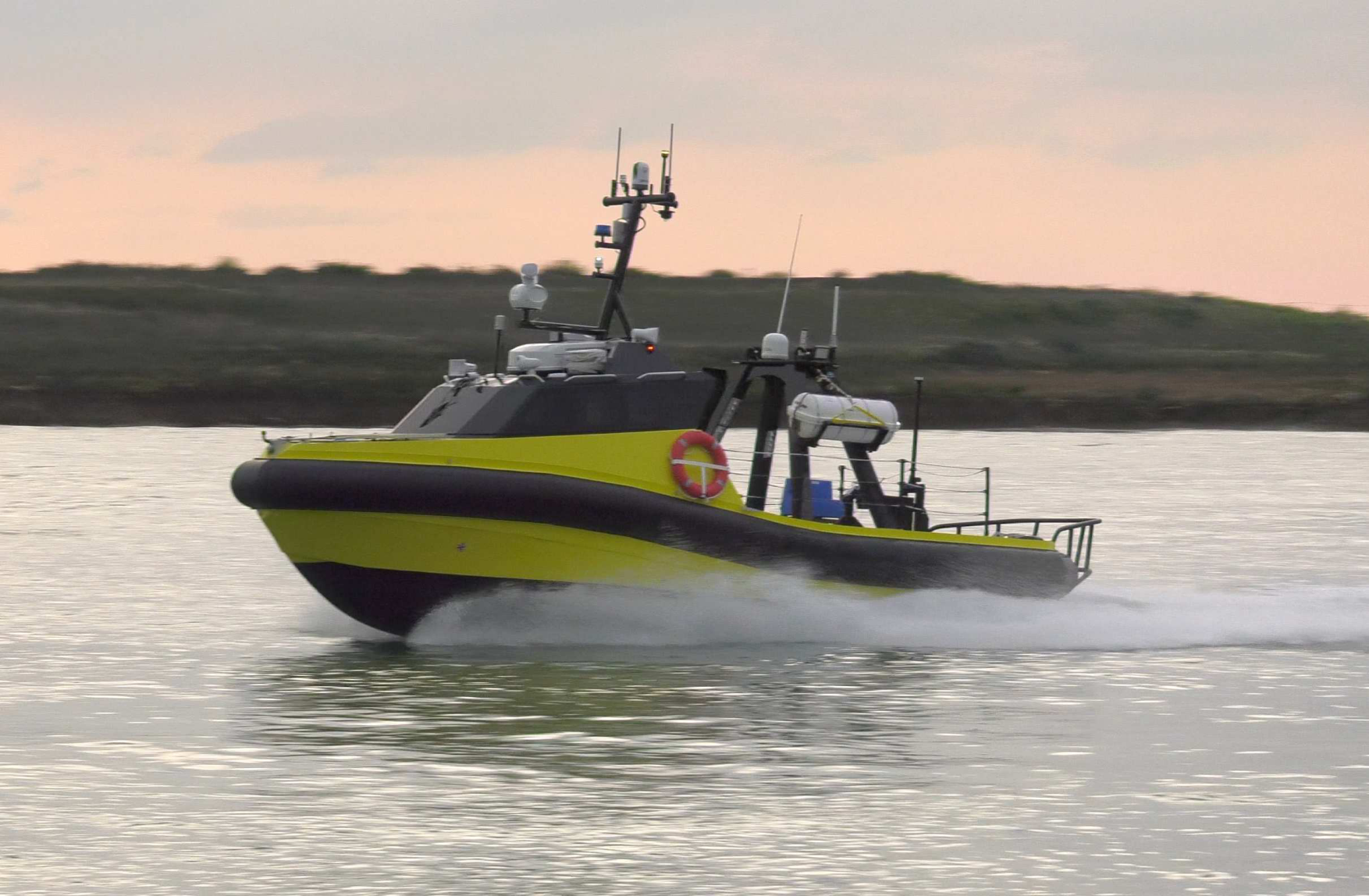}
    \includegraphics[width=0.35\textwidth]{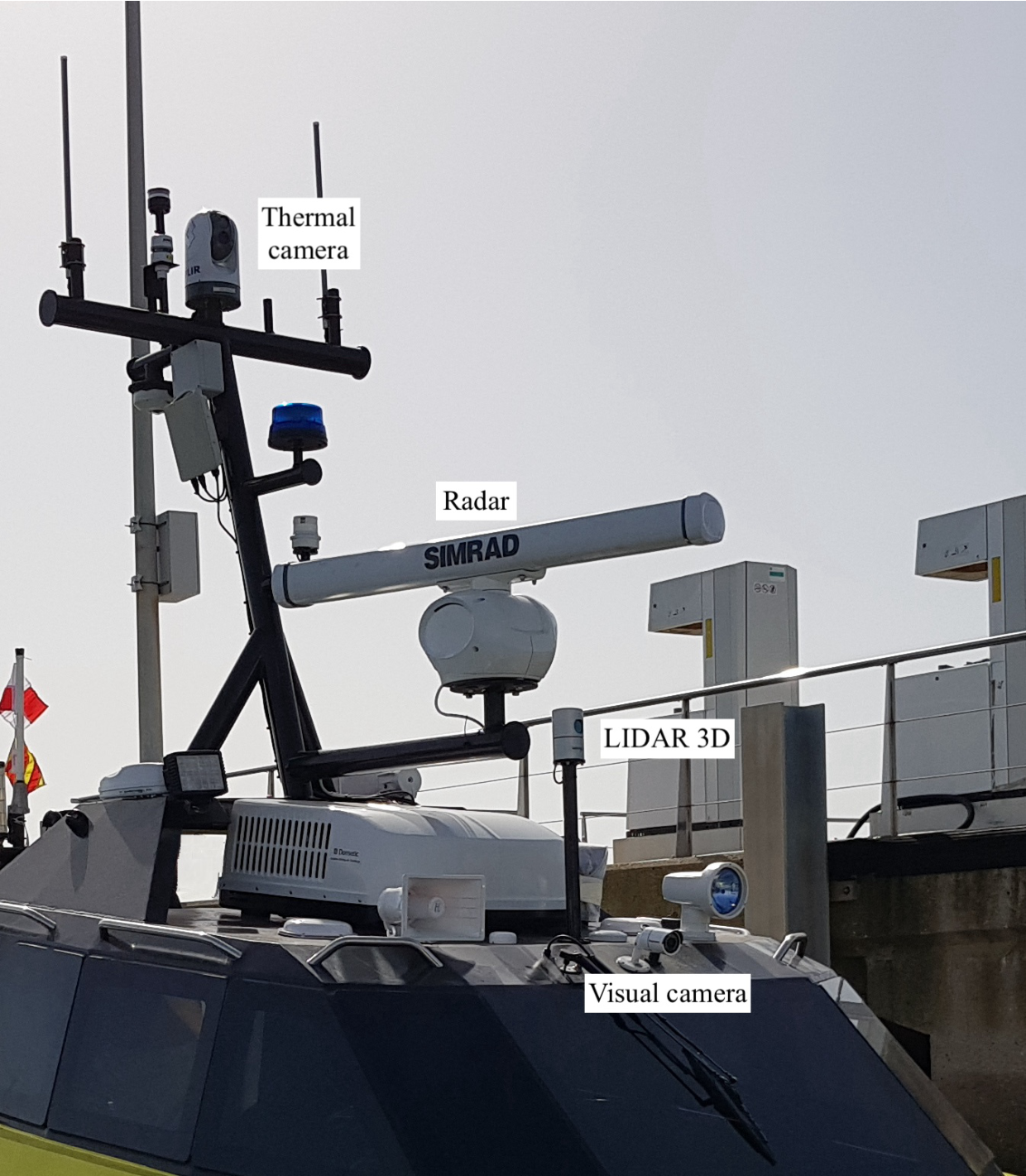}
    \caption{USV Vendaval.}
    \label{fig:USV}
\end{figure}

This article proposes a methodology for learning and controlling the autonomous guidance of a USV in a port environment. This includes both maneuvering in and out of the harbour as well as moving inside it. In a conventional manned vessel, the pilot, based on his experience, knowledge of the port, and applicable regulations, is in charge of tracing the correct trajectories taking into account the topology of the port and maritime signaling (buoys, beacons, and marks). Thus, for example, when a vessel arrives or departs a harbour it must follow channels delineated by the \emph{aid-to-navigation} devices defined by the IALA (International Association of Marine Aids to Navigation and Lighthouse Authorities) \cite{IALA}. Once inside the port, the pilot is in charge of avoiding shallow draft areas or other hazards, in addition to applying specific regulations that commercial ports may have.

\begin{figure}[htbp]
    \centering \includegraphics[width=0.9\textwidth]{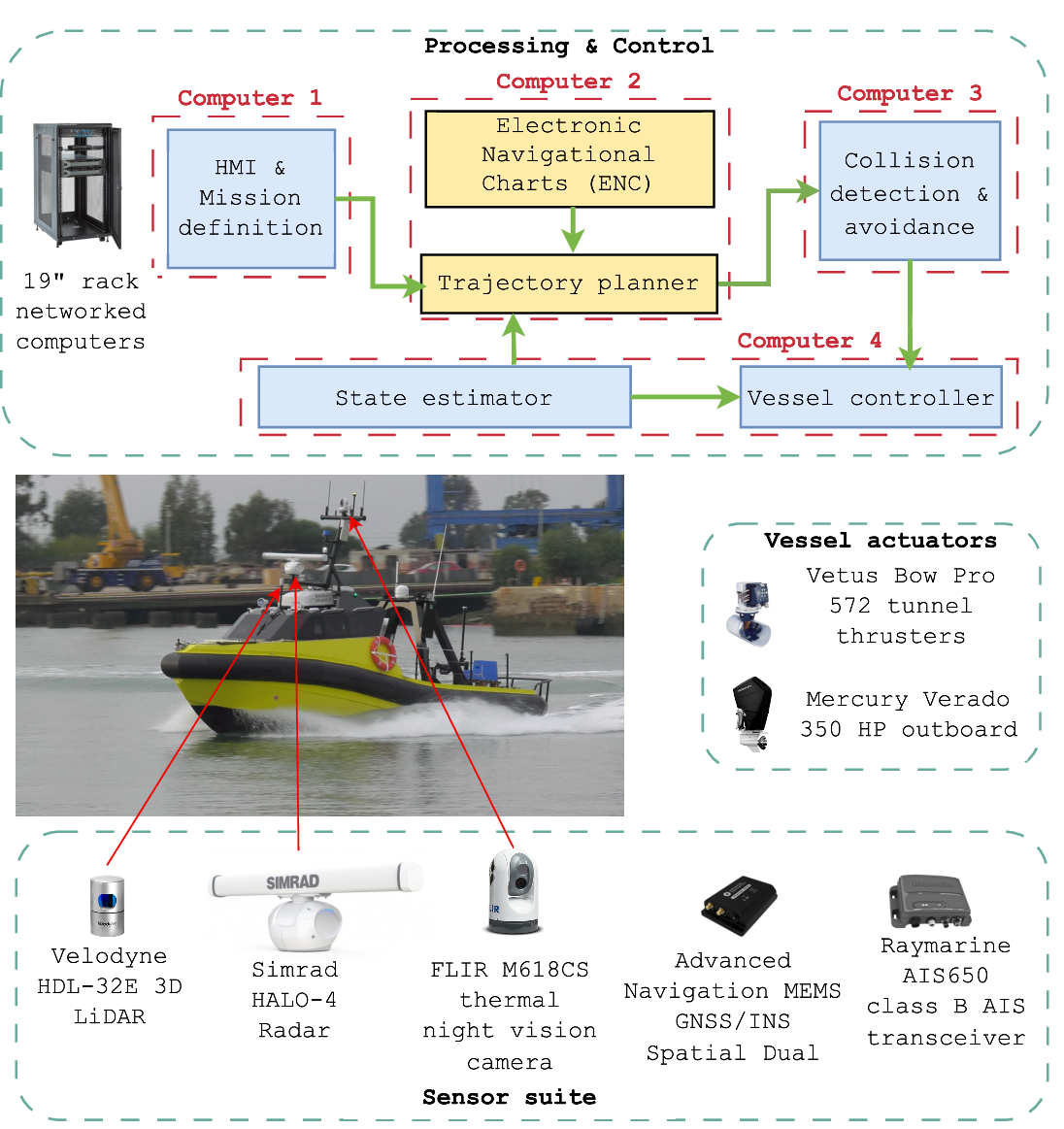}
    \caption{Vendaval USV functional architecture.}
    \label{fig:functional}
\end{figure}

In particular, the methodology presented in this paper is applied to the Vendaval USV to navigate at the port of Ceuta. Vendaval (see Fig. \ref{fig:USV} left) is a 10m length boat---owned by \href{https://www.navantia.es/en/}{Navantia} shipbuilding company---that can be operated as a manned, teleoperated or fully autonomous vessel. It is equipped with a Simrad HALO-4 marine radar, a Velodyne HDL-32 3D LIDAR, and a FLIR M618CS thermal night vision camera on top of its cabin as shown in Fig. \ref{fig:USV} right. The sensor suite also includes a Raymarine AIS650 class B AIS transponder, an Advanced Navigation Spatial Dual GNSS/INS receiver, and an EchoPilot 3D Forward Looking Sonar (FLS). Four networked computers are in charge of processing sensor data, executing navigation algorithms, communicating with the teleoperation/monitoring ground station, running a human-machine interface (HMI), and storing an electronically navigational chart (ENC). The boat is powered by a 350 HP Mercury Verado outboard engine and is maneuvered using two tunnel thrusters (see details in Fig. \ref{fig:functional}).

The USV autonomous navigation is based on the functional architecture also depicted at the top of Fig. \ref{fig:functional}, that includes the following subsystems: 

\begin{description}
    \item {\fontfamily{pcr}\selectfont Mission definition}: it allows to create a list of target waypoints, their tolerance, and the navigation velocities between them. It also allows to define restricted or forbidden areas.
    \item {\fontfamily{pcr}\selectfont Trajectory planner}: it provides a feasible trajectory that allows to reach the target waypoints while fulfilling all navigation constraints. These constraints are obtained from the {\fontfamily{pcr}\selectfont Electronic Navigational Charts (ENC)} server and the {\fontfamily{pcr}\selectfont Mission definition} subsystem.
    \item {\fontfamily{pcr}\selectfont State estimator}: it gathers data from the GNSS/INS sensor to estimate the vessel state (position, velocity, heading,  pitch, and roll).
    \item {\fontfamily{pcr}\selectfont Collision detection and avoidance}: it performs a data fusion algorithm combining proximity sensor data (radar, 3D LIDAR) with position information of surrounding boats obtained from the AIS receiver. This provides situational awareness information through a list of detected obstacles and an occupancy grid. If a potential collision is estimated, it provides a feasible maneuver and comes back to track the waypoints as soon as the collision risk disappears. Otherwise, reference signals are generated in order to follow the initial planned trajectory.
    
    \item {\fontfamily{pcr}\selectfont Vessel controller}: it calculates the control input needed in the vessel actuators using the vessel state, and reference signals from the {\fontfamily{pcr}\selectfont Collision detection and avoidance} subsystem.  
\end{description}

\begin{figure}[htbp]
    \centering \includegraphics[width=0.65\textwidth]{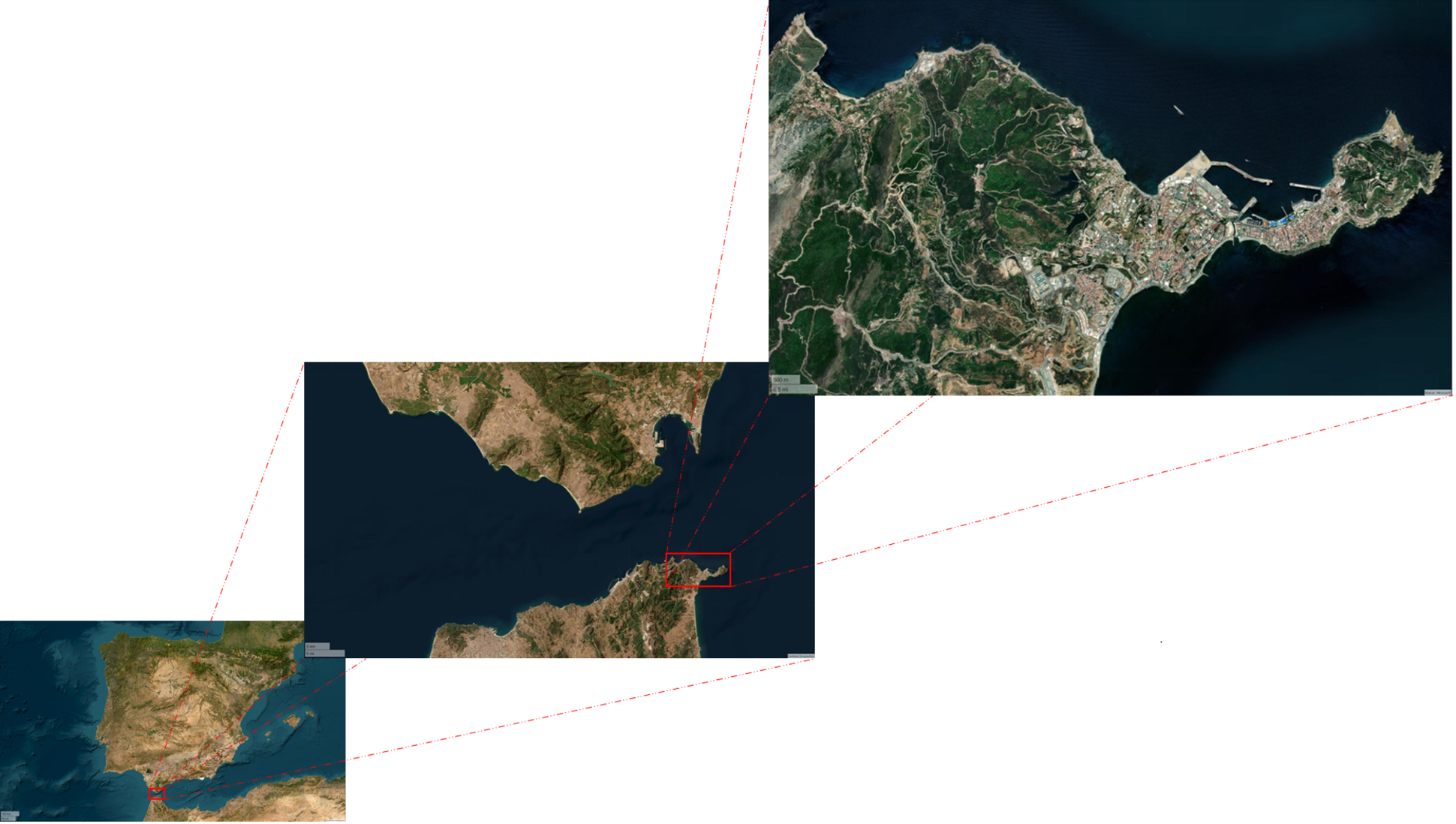}
    \caption{Map of the zone.}
    \label{fig:map2}
\end{figure}

\begin{figure}[htbp]
    \centering \includegraphics[width=0.90\textwidth]{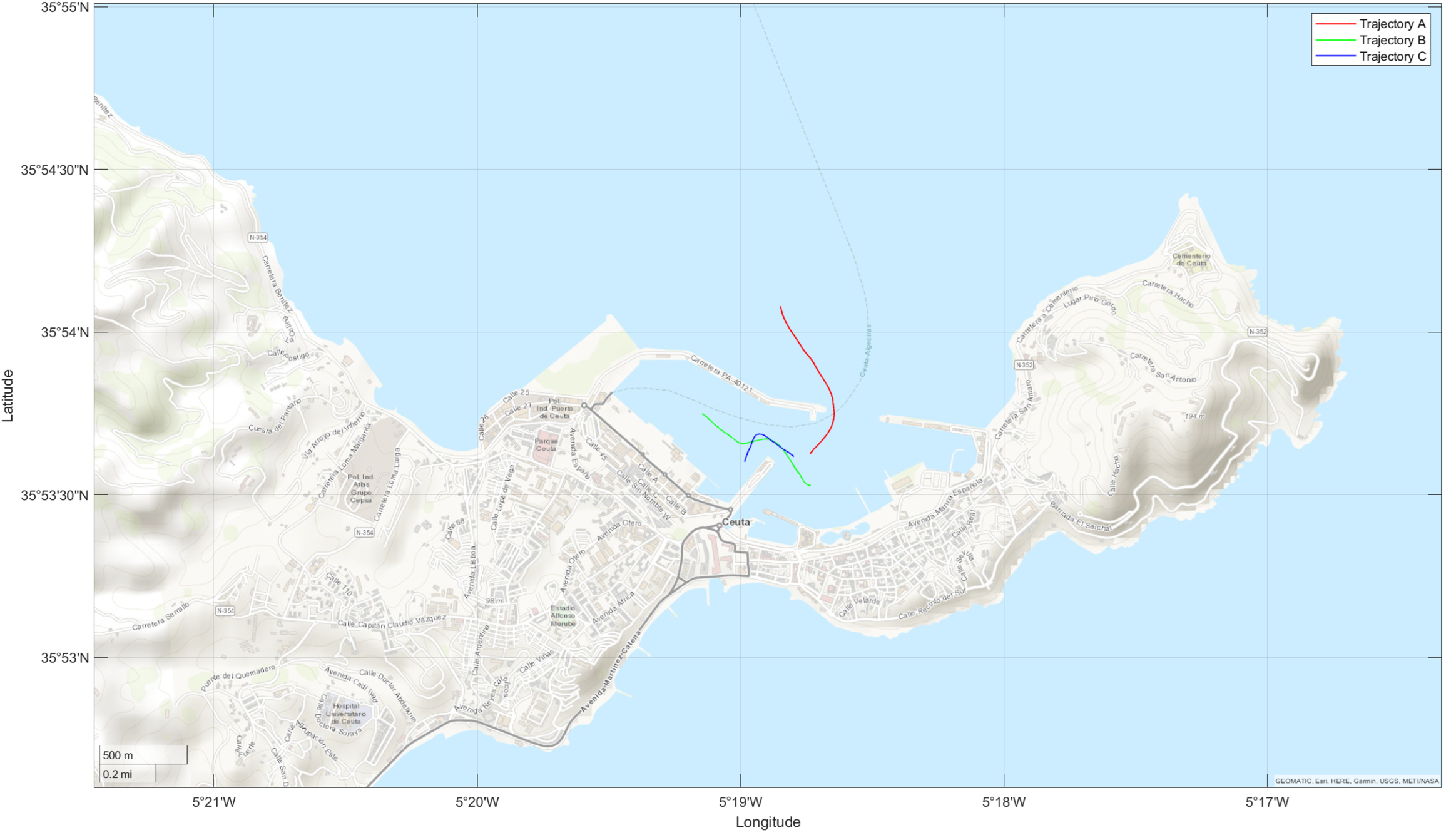}
    \caption{Trajectories at the zone.}
    \label{fig:zone}
\end{figure}

Ceuta is a Spanish city in North Africa (see Fig. \ref{fig:map2}). Its port is located on the northern shores of Morocco's coast, at the Mediterranean entrance to the Strait of Gibraltar. The port has two breakwaters (see Fig. \ref{fig:zone}) that measure 1500m and 500m long, known as the Poniente and Levante docks respectively. Inside the harbour, the control tower is located on the Spain quay (the main quay perpendicular to the coast). 

Fig. \ref{fig:zone} shows three types of standard trajectories. The USV sailing to enter the harbour (marked in red), the USV navigating from Levante to Poniente dock area (marked in green) and the USV moving around the control tower (marked in blue). 

%Therefore, the goal of the proposed methodology is to replace the Trajectory planner subsystem and the \textit{ENC} server (both marked in yellow in Fig. \ref{fig:functional}), so that, once the Vendaval USV is provided an initial position and a target position, a reference trajectory is generated that complies with all navigation constraints.

In brief, the goal of the proposed methodology is to replace the {\fontfamily{pcr}\selectfont Trajectory planner} and the {\fontfamily{pcr}\selectfont ENC} subsystems (marked in yellow in Fig. \ref{fig:functional}), so that, 
%once the Vendaval USV is provided an initial position and a target position, a reference trajectory is generated that complies with all port constraints.
once the pilot of the Vendaval provided a set of \emph{in situ} demonstrations, from initial to target position that comply with all port constraints and regulations, then: i) the trajectories will be recorded and subsequently learned and ii) the USV will be ready for autonomous navigation with controlled guarantees.

\section{Learning \& Control}

A learning of trajectories in a maritime environment presents significant challenges regarding to environmental conditions as ocean currents, waves and wind. Moreover, should the system learns a trajectory with specific conditions, analogous conditions are not guaranteed to be presented once the system reproduce autonomously the trajectory. Therefore, our approach does not only learn from the pilot-demonstrations, but guarantee system stability as well, in specific terms. Let us consider the continuous-time and smooth guidance control system defined as follows 
\begin{equation} \label{eq:sysu}
v = \hat{f}(x) + \hat u(x) + \eta(t),
\end{equation}
where $v$ stands for the velocity, $\hat{f}$ is a nonlinear estimate of the USV dynamics, $\hat{u}$ the input estimate and $\eta$ a bounded additive term accounting for noise in measurements and external perturbations, as e.g. those coming from maritime currents. Finally, the data-driven control system from \eqref{eq:sysu} comes from a sampled dataset denoted as $\mathcal{D}:=\{x^{m,n}, v^{m,n}\}$, which encompasses position $x \in \mathbb{R}^d$ and velocity $v \in \mathbb{R}^d$ vectors of dimension $d$, $m=1,...,M$ demonstrations and $n=1,...,N$ data-points each demonstration.

\bigskip
\noindent{\bf Problem statement.} Given a demonstration dataset $\mathcal{D}$ of port trajectories executed by an authorized pilot then, to learn the unknown dynamics of the USV $\hat f$ of \eqref{eq:sysu}, and guide it toward the target with appropriate control $\hat u$. For the sake of clarity, the target is defined as the origin $x=0$, without any loss of generality since a different target can be reproduced with a shift from it. Moreover, (robust) stability must be guaranteed while learning and control during the whole maneuver, because environmental conditions are inevitably present, collected in $\eta$ of \eqref{eq:sysu}.

\subsection{Fundamentals of GMM and GMR}

The underlying idea is to identify a nonlinear model like the one presented in the USV trajectories. Position and velocity for each demonstration and data-point are denoted as $[x^{m,n},v^{m,n}]$. Stacking them in matrix form with the demonstrations in rows as follows 

\begin{equation*}
	x= 
	\begin{bmatrix}
	x^{1,1} & x^{1,2} & \ldots & x^{1,n}\\
	x^{2,1} & x^{2,2} & \ldots & x^{2,n}\\
	\vdots & \vdots & \ddots & \vdots\\
	x^{m,1} & x^{m,2} & \ldots & x^{m,n}\\
	\end{bmatrix} \textrm{,} \quad
	v= 
	\begin{bmatrix}
	v^{1,1} & v^{1,2} & \ldots & v^{1,n}\\
	v^{2,1} & v^{2,2} & \ldots & v^{2,n}\\
	\vdots & \vdots & \ddots & \vdots\\
	v^{m,1} & v^{m,2} & \ldots & v^{m,n}\\
	\end{bmatrix} \textrm{.}	
\end{equation*}

A Gaussian Mixture Model (GMM) is employed to address such a problem. GMM is an unsupervised learning method based on the weighted sum of probability density function $P(x^{m,n},v^{m,n}|{\bm \theta_k})$ of a finite set of Gaussian kernels $K$, which is defined as follows

\begin{equation*}
P(x^{m,n},v^{m,n}|{\bm \theta_k}) = \sum_{k=1}^{K}\pi_k \mathcal{P}({x^{m,n},v^{m.n}}|k)
\
    \begin{dcases}
        k \in 1,\ldots,K,
        \\
        n \in 1,\ldots,N,
        \\
        m \in 1,\ldots,M.
    \end{dcases}
\end{equation*}

\noindent where ${\bm \theta_k}:=\{\pi_k,\mu_{k},\Sigma_{k}\}$ with $k=1,\ldots,K$ are the parameters that describe each and every Gaussian distribution. Hence, GMM is a probabilistic model used to cluster the data. 

A practical way for finding the optimal parameters ${\bm \theta_k}$ is through the Expectation-Maximization (EM) algorithm \cite{r22}. EM is described by 4 steps: Parameter initialization, E-step, M-step and log likelihood evaluation (see \cite{Bishop} for more detail). The first step is typically accomplished by the K-means algorithm \cite{Bishop}; the second step computes the posterior probability $\gamma_{k}(x,v) \in \mathbb{R}_{[0,1]}$ once $[x^{m,n},v^{m,n}]$ is observed as follows
\begin{equation*}
    \gamma_{k}(x,v):=\frac{\pi_k \mathcal{P}(x^{m,n},v^{m,n}|k)}{\sum_{i=1}^{K}\pi_i \mathcal{P}(x^{m,n},v^{m,n}|i)},
\end{equation*}
\noindent where $\pi_k \in \mathbb{R}^{k}$ is the prior probability and $\mathcal{P}(x^{m,n},v^{m,n}|k)$ is the probability that a data-point $[x^{m,n},v^{m,n}]$ belongs to a Gaussian kernel $k$, which is defined as a Gaussian normal distribution 
\begin{equation*}
\begin{split}
\mathcal{P}(x^{m,n},v^{m,n}|k) & = \mathcal{N}(x^{m,n},v^{m,n}; \mu_{k}, \Sigma_{k}) \\ & = {\left({(2\pi)}^{2d}{\rm det}({\Sigma}_k)\right)}^{-\frac{1}{2}}\exp{\left(-\frac{1}{2} \Bigl( [x^{m,n},v^{m,n}]-\mu_k \Bigr)^T {(\Sigma_k)^{-1}} \Bigl( [x^{m,n},v^{m,n}]-\mu_k \Bigr) \right)},
\end{split}
\end{equation*}
\noindent the third step updates the parameters ${\bm \theta_k}$ iteratively as follows
\begin{align*}
\pi_k^{new} &= \frac{\Gamma_{k}}{MN}, \\
\mu_k^{new} &= \frac{1}{\Gamma_{k}}	\sum_{m=1}^{M}\sum_{n=1}^{N} \gamma_{k}(x,v)[x^{m,n},v^{m,n}], \\
\Sigma_k^{new} &= \frac{1}{\Gamma_{k}} \sum_{m=1}^{M}\sum_{n=1}^{N} \gamma_{k}(x,v)\Bigl( [x^{m,n},v^{m,n}]-\mu_k^{new} \Bigr)^T \Bigl( [x^{m,n},v^{m,n}]-\mu_k^{new} \Bigr),
\end{align*}

\noindent where $\Gamma_{k}:=\sum_{m=1}^{M}\sum_{n=1}^{N}\gamma_{k}(x,v)$. EM seeks for maximizing the likelihood with respect to ${\bm \theta_k}$ so that, in the fourth step it evaluates such function as follows  
\begin{equation*}
\log{P(x^{m,n},v^{m,n}|{\bm \theta_k})} = \sum_{m=1}^{M}\sum_{n=1}^{N}\log{\sum_{k=1}^{K}\pi_k \mathcal{P}({x^{m,n},v^{m.n}}|k)}.
\end{equation*}

Thus, the parameters ${\bm \theta_k}$ are initialized and optimized from the dataset $\mathcal{D}$. The resulting mean vector $\mu_{k} \in \mathbb{R}^{2d}$ and the covariance matrix $\Sigma_k \in \mathbb{R}^{2d \text{x} 2d}$ for a $k$ Gaussian are defined as
\begin{equation}
    \mu_k = 
    \begin{bmatrix}
        \mu_{k}^{x}
        \\
        \mu_{k}^{v}
    \end{bmatrix} \textrm{,} \quad
    \Sigma_k = 
    \begin{bmatrix}
        \Sigma_{k}^{x} & \Sigma_{k}^{xv}
        \\
        \Sigma_{k}^{vx} & \Sigma_{k}^{v}
   \end{bmatrix},
\end{equation}

One of the GMM benefits as described in \cite{r23} is that there is no difference between outputs and inputs; hence, we have selected position as an input and velocity as an output. Once GMM have found the optimal parameters for ${\bm \theta_k}$, it is feasible to estimate a velocity $\hat v$ through Gaussian Process Regression (GMR), which is defined through the weighted conditional mean as
\begin{equation} \label{eq:fest}
    \hat{v} = {\sum_{k=1}^{K} \gamma_{k}(x) \left(\mu_{k}^{v} + \Sigma_{k}^{vx}(\Sigma_{k}^{x})^{-1}(x-\mu_{k}^{x})\right)}.
\end{equation}
Notice that the influence measurement of different Gaussians is defined as a nonlinear weighting term $\gamma_{k}(x) \in \mathbb{R}_{[0,1]}$ described in \cite{Bishop}, which is given by
\begin{equation*}
    \gamma_{k}(x):=\frac{\pi_k \mathcal{P}(x^{m,n}|k)}{\sum_{i=1}^{K}\pi_i \mathcal{P}(x^{m,n}|i)},
\end{equation*}
\noindent where $\mathcal{P}(x^{m,n}|k)$ is the probability of $x^{m,n}$ given $k$ and it is defined as
\begin{equation*}
    \mathcal{P}(x^{m,n}|k) = \left(2\pi\Sigma_k^{x}\right)^{-1/2}\exp{\left(-\frac{({x^{m,n}}-\mu_{k}^{x})^2}{2\Sigma_k^{x}}\right)}.
\end{equation*} 

The resulting nonlinear velocity $\hat{v}$ is capable to estimate a huge diversity of trajectories due to the nonlinear weighting term $\gamma_{k}(x)$. Note that $\hat{v}$ can be understood as a nonlinear weighted sum of linear dynamical systems.   

\begin{figure}[htbp]
    \centering 
    %\captionsetup{justification=centering}
    %\includegraphics[width=0.55\textwidth,height=4.5cm]
    \includegraphics[width=0.75\textwidth]{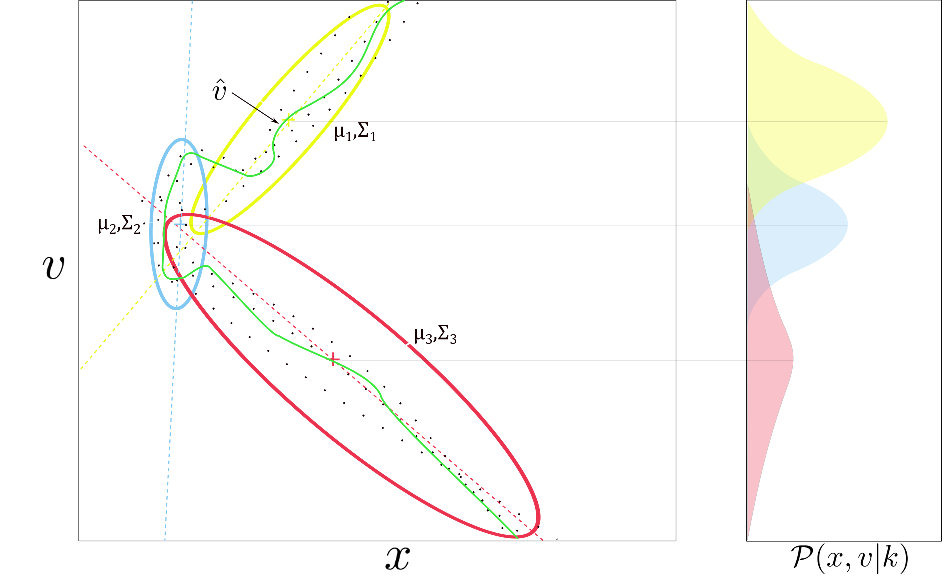}
    \caption{\small Adjustment of three Gaussian kernels for a 1D system}
    \label{fig:GMMGMR-explanation}
\end{figure}

\begin{figure}[htbp]
    \centering 
    %\captionsetup{justification=centering}
    %\includegraphics[width=0.55\textwidth,height=4.5cm]
    \includegraphics[width=0.7\textwidth]{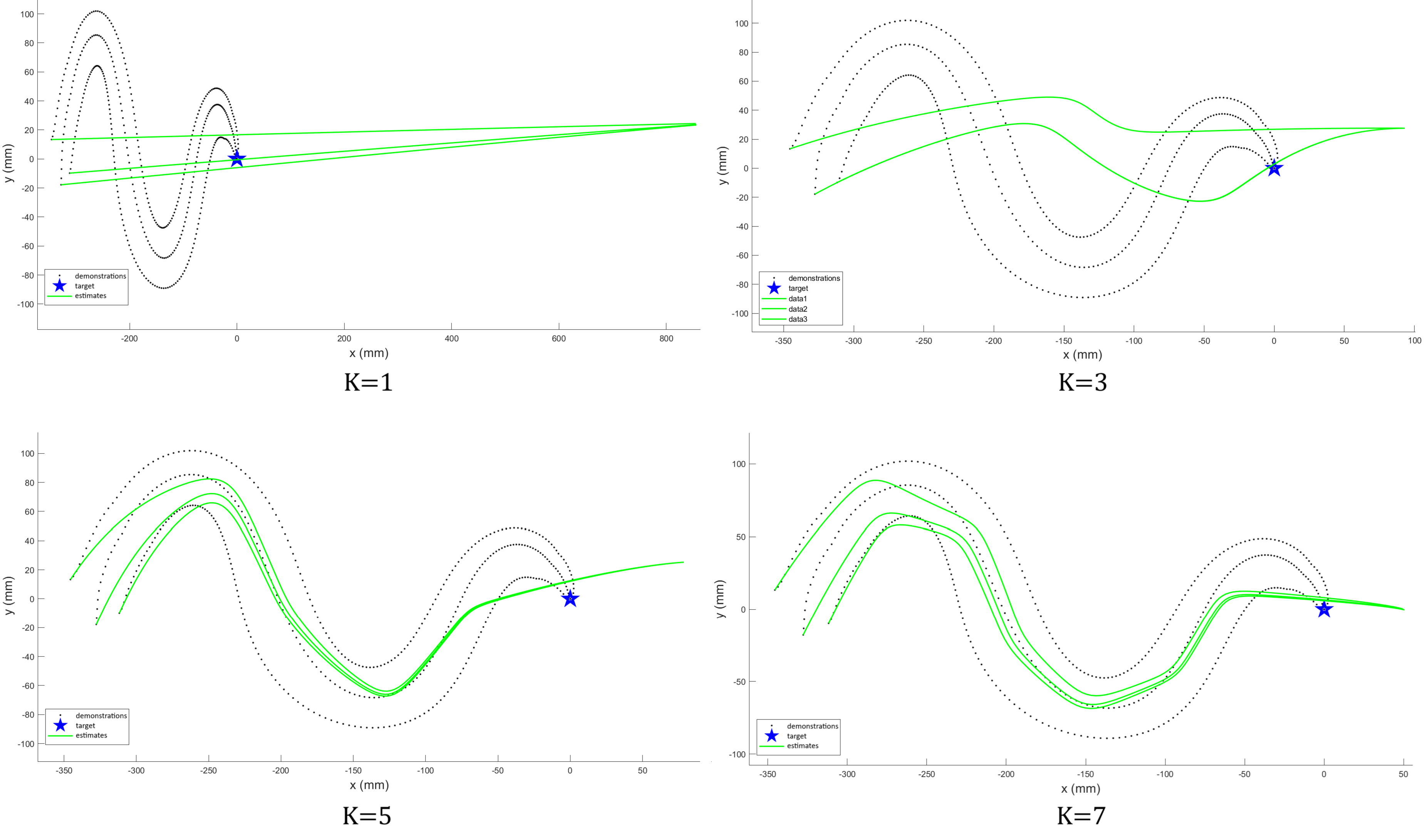}
    \caption{\small GMM/GMR with different Gaussian Kernels}
    \label{fig:GMMGMR-example}
\end{figure}

The use of GMM/GMR allows nonlinear systems to be estimated. Its mechanism is shown in Fig. \ref{fig:GMMGMR-explanation}, where three Gaussian kernels are combined to estimate a general trajectory for a 1D system. However, depending on the trajectory complexity, it requires more or less number of Gaussian kernels to reproduce an acceptable estimate and eventually, converge near to a target. A toy example is used in Fig. \ref{fig:GMMGMR-example} to depict the effect of Gaussian kernels over the estimate. Note that the greater the number of kernels the better the estimate; furthermore, no estimate is able to converge to the target (star). Hence, the following section presents a control law to achieve a good estimate with stability guarantees, using a reduced number of $K$.

\subsection{Data-driven Control}
%\subsection{Control Lyapunov Function}

As it was stated before, GMM/GMR is not sufficient for guaranteeing convergence to the target. 
This is especially critical in this marine application, as environmental disturbance, such as ocean currents, can cause drifts in estimates or even instabilities during executions and, eventually, hazard situations.
The proposed controller is data-driven and based on the Sontag's Universal formula; furthermore, it has been first proposed in \cite{BecAco24} in the general context of the discovery of nonlinear dynamics. The good learning performance in noisy environments and the disturbance rejection capabilities make it suitable for this marine application.
The controller is based on a Control Lyapunov Function (CLF), concept that was first introduced in \cite{Arstein,Sontag} and first used in a marine application on dynamic positioning of ships in \cite{Acosta} in a different framework of continuous-time constrained control and without learning. In essence, the CLF, namely $V(x)$, must be differentiable, positive definite $V(x) > 0$, $V(0) = 0$ and satisfy the inequality given by       
\begin{equation} \label{clf-opt}
\inf_{\hat u\in \mathcal{U}} \ \{\nabla_x^\top V(x) (\hat{f}({x}) + \hat u(x))\} < 0,
\end{equation}
thus, guaranteeing the existence of such $\hat u$ that enforces the decrease of $V(x)$ along the trajectories of \eqref{eq:sysu}. For that reason, it is also called energy-like function. For those that are not familiar with it, notice that it can be seen as the solution of a constrained optimization problem on $\hat u$. In this work, we define a Weighted Sum of Asymmetric Quadratic Function (WSAQF) to construct a CLF satisfying \eqref{clf-opt}, similarly to \cite{BecAco24} and references therein. Thus, the CLF candidate is given by
\begin{equation} \label{eq:V1}
V({ x}) = { x}^{\top} P_{0} { x} + \sum_{l=1}^{L} \sign_{+}(\sigma_{l}) \ \sigma_{l}({ x})^{2},
\end{equation}
where $\sigma_{l}({ x}):={ x}^{\top} P_{l} ({ x - \mu}_{l})$ and $P_l \in \mathbb{R}^{d\,\mathrm{x}\,d}$ are positive definite matrices; $\mu_l\in \mathbb{R}^{d}$ are mean vectors; $L$ is the number of asymmetric quadratic functions; and $\sign_{+}(\cdot)$ denotes a coefficient defined as 0 when its argument is negative and 1 elsewhere. Notice that \eqref{eq:V1} is globally positive definite and convex by construction, which can be seen from the fact that $\nabla_{x} V(0)=0$ and $x^{\top} \nabla_{x} V(x)>0$ for all $x\neq 0$ (see Lemma 1 of \cite{BecAco24} for a detailed proof).
%
%Certainly, by direct calculation we get
%%
%\begin{align*}
%x^\top \nabla_{x} V({x}) &= x^\top (P_{0}+P_{0}^{\top}) x + x^\top \sum_{l=1}^{L} 2\sign_{+}(\sigma_{l}) \sigma_{l}({ x}) (P_{l} ( x - \mu_{l})+P_{l}^{\top}x) \\
%%&= x^\top (P_{0}+P_{0}^{\top}) x + \sum_{l=1}^{L} 2\sign_{+}(\sigma_{l})  \sigma_{l}({ x}) x^{\top } P_{l} ( x - \mu_{l}) + 2\sign_{+}(\sigma_{l})  \sigma_{l}({ x}) x^{\top } P_{l}^{\top} x \\
%&= x^\top (P_{0}+P_{0}^{\top}) x + \sum_{l=1}^{L} \Big( 2 \sign_{+}(\sigma_{l})  \sigma_{l}({ x})^{2} + \sign_{+}(\sigma_{l}) \sigma_{l}({ x}) x^{\top } (P_{l} + P_{l}^{\top}) x \Big) \geq 0.
%\end{align*}%
%and $\sign_{+}(\sigma_{l}) \sigma_{l}\geq 0$
%
Once the CLF is defined, the controller is defined through a simpler version of the Sontag's Universal formula, which is given by the following continuous-time state feedback
\begin{equation}\label{eq:uopt}
    %\scalemath{0.75}
    {\hat u(x):=} 
    %\scalemath{0.75}
    {\begin{dcases}
        - (a(x) + \rho(| x|)) {\overrightarrow{b}}(x), & a(x) + \rho(|x|) > 0, \\
        0, & a(x) + \rho(|x|) \leq 0,
    \end{dcases}} 
\end{equation}
where $a(x):=\nabla_x V(x)^{\top} \hat{f}(x)$, ${\overrightarrow{b}(x)}:= b(x)^\top/|b(x)|^{2}$ with $ b(x):=\nabla_x V(x)^{\top}$. The gradient of $V$ can be obtained easily by direct calculation from \eqref{eq:V1} resulting in
\begin{equation} \label{eq:Vnabla}
\nabla_{x} V({x}) = (P_{0}+P_{0}^{\top}) x + \sum_{l=1}^{L} 2\sign_{+}(\sigma_{l}) \sigma_{l}({ x}) (P_{l} ( x - \mu_{l})+P_{l}^{\top}x).
\end{equation}
The positive function $\rho(|x|)$ completes the definition of the explicit controller. For that,  Sontag proposed $\rho(|x|) := \rho_0 \sqrt{a(x)^{2}+|b(x)|^{4}}$ with $\rho_0>0$, that is the solution of the optimal control problem and hence it guarantees convergence to the target.

\subsection{Optimization-based learning problem}

Learning of complex trajectories is performed through an optimization problem. Thereby, GMM/GMR and the CLF are concatenated to yield robust and asymptotic-like stable estimates from the data. The possible solutions for such a problem determine the optimal parameters $\bm \theta_{\bm k}$ for GMM/GMR and the optimal parameters ($P_l$, $\mu_l$) for the CLF. Let $\bm \theta:= \{\bm \theta_{\bm k}, P_l, \mu_l\}$ be the whole set of parameters. The optimization problem initializes with a trajectory estimate performed by GMM/GMR and with identity matrices and null vectors for the energy function. 

An eager learning algorithm is proposed in order to determine the optimal parameters for $\bm \theta$. However, note that the number of Gaussian Kernels $K$ and asymmetric quadratic functions $L$ employed to model the nonlinear system complexity defines the number of parameters to be calculated, i.e. the more-complex trajectories the greater number of optimal parameters and difficulty of learning.
%more-complex trajectories produce greater number of optimal parameters as well as greater difficulty in learning them.

The objective function proposed seeks to minimize the error between the measured velocity $v$ and the estimate velocity $\hat{v}$. The optimization problem can be formalized as follows
\begin{mini}|s| %<b>
{\bm \theta}{J(\bm \theta) :=\frac{1}{2 N M} \sum_{n=1}^{N} \sum_{m=1}^{M} | {v}^{n,m}-\hat{v}^{n,m} |^{2}}
{\label{J}}{}
\addConstraint{\Sigma_{k=1}^{K} \pi^{k}}{=1;}{0 < \pi^{k}}{<1}
\addConstraint{\Sigma_{k}}{\succ 0,\quad}{k=1,\ldots,K}
\addConstraint{P_l}{\succ 0,\quad }{l=0,\ldots,L.}
\end{mini}
As it can be seen, it is a non-convex problem and hence, solvers can not guarantee to find a global solution, but a local minima can be found through methods such as the Interior Point algorithm \cite{Wright}. Thus, this approach is employed to solve the constrained optimization problem \eqref{J}, which is summarized in Algorithm~\ref{alg}.

\begin{algorithm}
	\caption{CLF-based Stable Estimator} 
	\hspace*{\algorithmicindent} \textbf{Input:} $[x^{m,n},v^{m,n}]$, $\rho_0$, $M,N,K$ and $L$
	\begin{algorithmic}[1]
	    \State Run GMM to compute initial parameters [$\pi_{k_0}$, $\mu_{k_0}$, $\Sigma_{k_0}$] for the while loop
	    \State Initialize $P_{l_0}$ and $\mu_{l_0}$
%	    \State Set a constraint for $P_L$ as a Positive Definite Matrix
	    \While {$J>$ \textbf{threshold} of \eqref{J} is not satisfied}
	    \State Estimate $\hat{{v}}^{m,n}$ from GMR $\bm \theta_{k}$ using \eqref{eq:fest}
	    \State Estimate $V(x^{m,n})$ from $P_l$ and $\mu_l$ using \eqref{eq:V1} 
	    \State Compute  $\dot{V}(x^{m,n},\hat{v}^{m,n})$ from \eqref{clf-opt},  \eqref{eq:uopt} and \eqref{eq:Vnabla}
	    \State Compute $u^{m,n}$ from \eqref{eq:uopt}
	    \Indent
	    	\If{$\dot{V}(x^{m,n},\hat{v}^{m,n})$ $>$ 0}
	     %\State $(\nabla_xV(x^{m,n}))^T(\hat{\dot{x}} + u^{m,n})$ from Eq. (9)
		\ $(\hat{{v}} + u^{m,n})$
	    \ElsIf {$\dot{V}(x^{m,n},\hat{v}^{m,n})$ $\leq$ 0}
	        %\State $(\nabla_xV(x^{m,n}))^T\hat{\dot{x}}$  from Eq. (9)
	    \ $\hat{{v}}$
	    \EndIf
	    \EndIndent
	    \State Minimize $|{v}^{m,n} - \hat{{v}}^{m,n}|^{2}$ of \eqref{J}
        \EndWhile 

	\end{algorithmic} 
	\hspace*{\algorithmicindent} \textbf{Output:} $\bm \theta^{*} =\{\pi_K^*, \mu_K^*, \Sigma_K^*, P_L^*, \mu_L^*\}$
    \label{alg}
\end{algorithm}

\section{Evaluation and Results}

The experiments have been carried out at the port of Ceuta with the Vendaval USV. The main features of this autonomous ship are listed in Table \ref{tab:1}. The USV can be manually piloted, teleoperated from a control spot or set to autonomously follow pre-programmed trajectories. In order to obtain relevant data from the maritime trajectory for the proposed learning method in this work, the USV is manually piloted by an expert user at the beginning of the experiments.    

\begin{table}[h]
\centering
\begin{tabular}{|c c|}
\hline
Parameters & Values  \\
\hline\hline
Length & 10 m \\
Beam & 3 m \\
Draught & {0.65 m} \\
Velocity & 25 knots (45 km/h) \\
Outboard engine & 350 hp \\
%Mass & \textcolor{red}{? kg} \\
\hline
\end{tabular}
\caption{USV Vendaval features}
\label{tab:1}
\end{table}

The pilot based on his expertise and the ECDIS (Electronic Chart Display and Information System) indications manually steers the USV from different but geographically close initial positions to at unique target. Data position and heading of the USV are measured and recorded while an expert user is piloting it through a specific trajectory. The same exercise is repeated three times in order to construct a set of demonstrations from it, but more importantly, to generalize the trajectory learning. The recorded data is adapted for the set of demonstrations as follows:

\begin{enumerate}
  \item The geographic coordinates are transformed to UTM (Universal Transverse Mercator). Such a transformation is carried out by ETRS89 (European Terrestrial Reference System 1989), which allocates coordinate axes to every zone of the terrestrial surface without regarding Earth curvature.
  \item The target of every demonstration is set to the origin of coordinates (0,0).
  \item The performed demonstrations over a real environment do not reach the same target due to the measurement error made by the pilot to get to the same position. Hence, the demonstrations are slightly corrected offline to converge to the same target, in order to `show' the desired target point to the learning algorithm. 
\end{enumerate}

%\begin{figure}[htbp]
    %\centering 
    %\includegraphics[width=0.55\textwidth]{SetDemonstrationsUSV.eps}
    %\caption{\small Set of Demonstrations for a trajectory of the USV}
    %\label{fig:setdem}
%\end{figure}

Once the set of demonstrations is constructed, the recorded data is used as an input for our method to get estimates for every trajectory. A good trajectory estimate as well as its convergence depends on tuning the number of Gaussian functions $K$, the number of asymmetric functions $L$, and the parameter $\rho_0$. On the one hand, the Gaussian and asymmetric functions are established according to the trajectory complexity to learn. On the other hand, the controller gain $\rho_0$ must be a small enough to guarantee the convergence toward the target point without affecting the estimate quality to such an extent. Greater $\rho_0$ values will prioritize the control law, forgetting the estimate fidelity.  

\subsection{Estimate of USV position}

\begin{figure}[htbp]
    \centering 
    \includegraphics[width=1\textwidth]{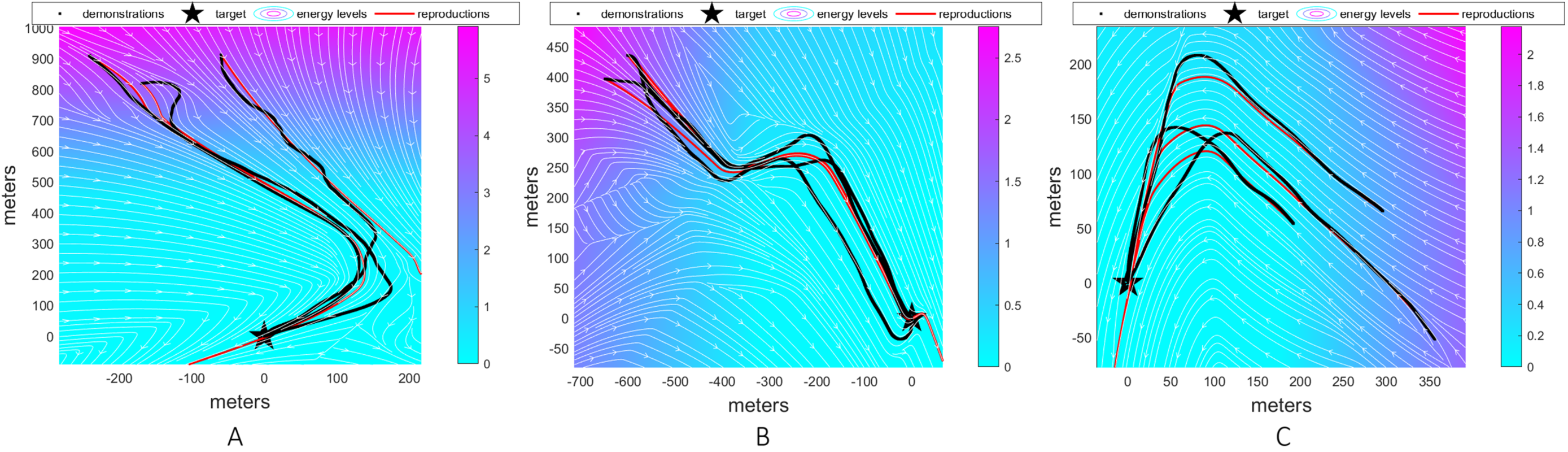}
    \caption{\small Estimates of USV position using GMR}
    \label{fig:GMR_2D}
\end{figure}

Our first set of experiments is conducted to estimate USV position. The pilot navigated the USV through different waterways at the port of Ceuta. The approximated distances of every trajectory are between 500 m and 1 km from north to south as well as west to east. Three different trajectories were performed in order to validate our method. The solid black lines stand for performed demonstrations by the pilot, whereas the solid red lines stand for estimates of such demonstrations. A learning method like GMM/GMR is a powerful tool to estimate data as presented in section 3, but its main drawback is to preserve stability conditions. The depicted estimates in Fig. \ref{fig:GMR_2D} do not converge toward the target and are susceptible to reproduce trajectories totally unstable, should a perturbation is presented. In these trajectories, a reduced number of Gaussians is used ($K=5$) because of the low-complexity nonlinearities and the number of asymmetric functions is $1$ (the same number of $L$ is used for the following experiments).

\begin{figure}[htbp]
    \centering 
    \includegraphics[width=1\textwidth]{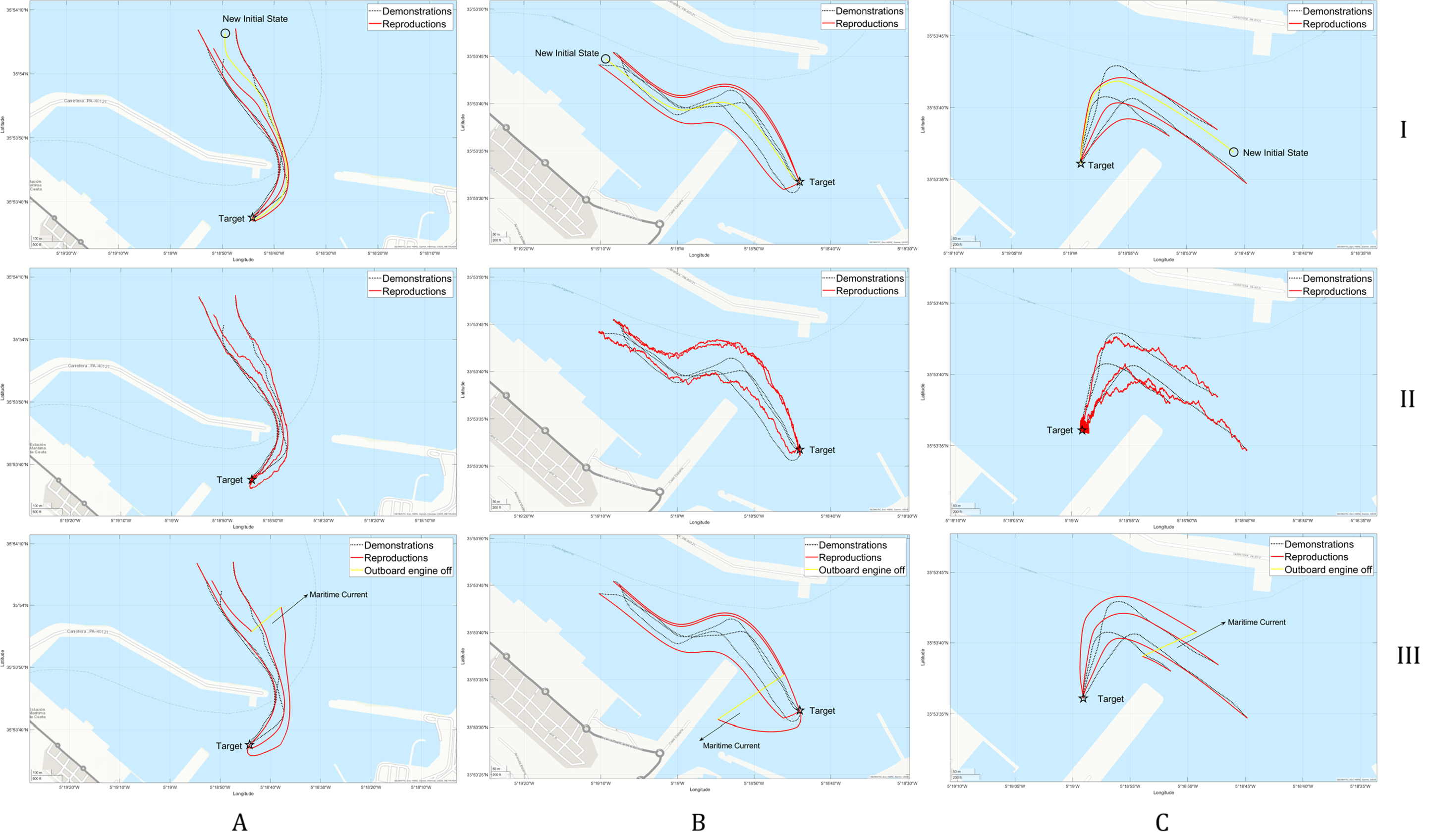}
    \caption{\small Estimates of USV position at the port of Ceuta. I) Non-disturbances II) Disturbances III) Localized perturbation}
    \label{fig:XY}
\end{figure}

Estimates can be occasionally affected by the nature of the environment (i.e. maritime currents), taking the USV to undesirable location. Nevertheless, it depends on the strength of the disturbance. In some cases, the effect of maritime currents over the USV can be neglected in calm waters (see Fig.\ref{fig:XY}-I). In some other cases, this effect can not be ignored; hence, the USV has been perturbed on purpose in order to validate the system robustness to track or even return to the trajectory. On the one hand, the USV behavior under constant disturbances is depicted in Fig. \ref{fig:XY}-II. The estimates keep replicating the real system behavior and the convergence is pretty close to the origin. None of them reach the target, though. On the other hand, the USV is able to recover from a big localized perturbation (in order to reproduce such perturbations, the outboard engine was turned off for a period of time, so that, the USV swept away in maritime currents) as depicted in Fig. \ref{fig:XY}-III. 
%Note that the control signal effect increases upon the perturbation is presented in order to force the system at the target (i.e. asymptotic stability). 
Note that the estimates start from the same position that the demonstrations do; however, a different position from those used by the pilot is employed to validate the learning model and its respective convergence to the target (see Fig.\ref{fig:XY}-I - yellow trajectory).  

\begin{figure}[htbp]
    \centering 
    \includegraphics[width=1\textwidth]{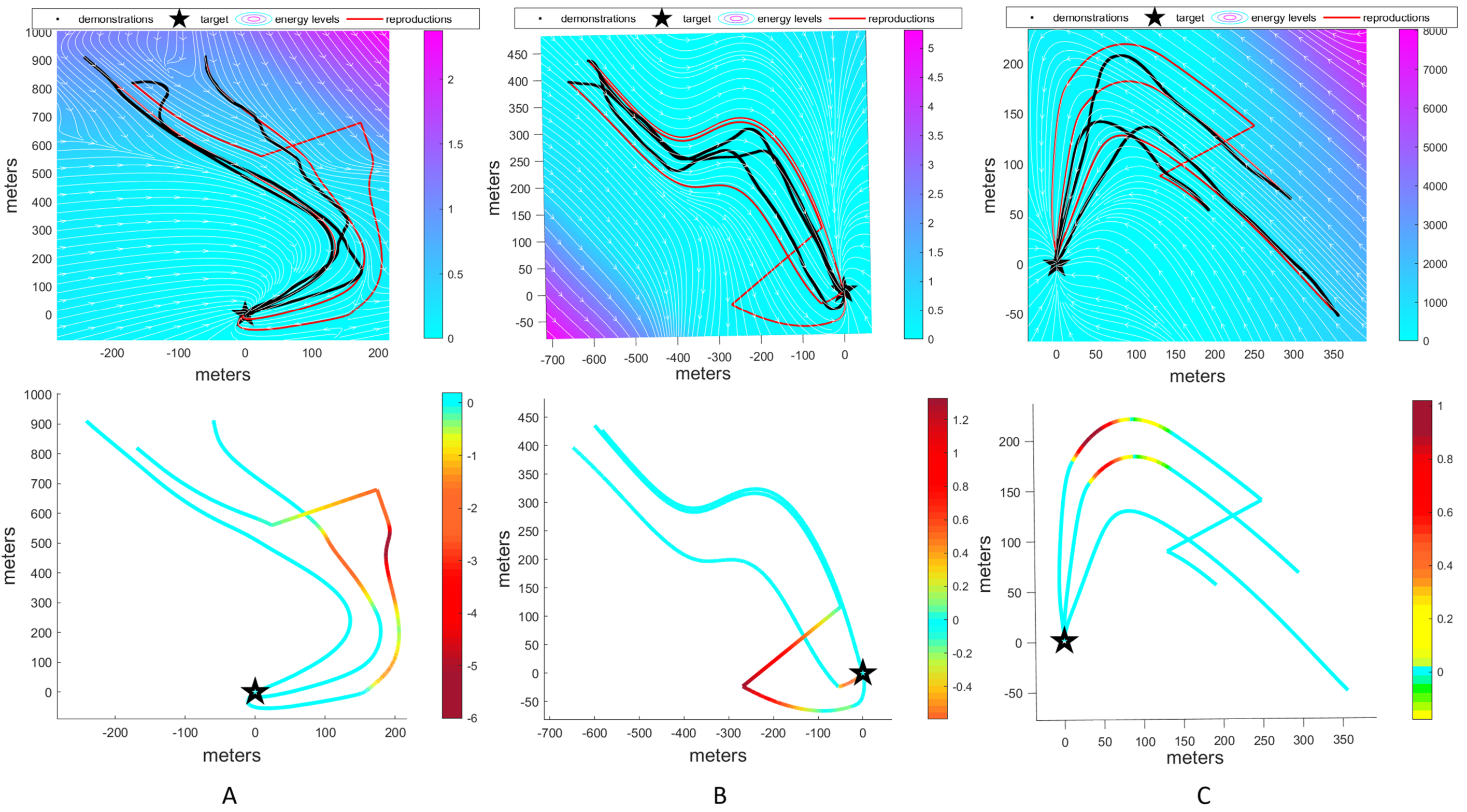}
    \caption{\small Control signal effect}
    \label{fig:Control}
\end{figure}

A more detailed analysis of the data-driven control is exposed in Fig.\ref{fig:Control}. The streamlines converge toward the target alongside a decreasing energy levels. Moreover, the proposed control law is able to correct the affected estimate by the localized perturbation. Hence, a minimal control effort is done as long as the estimates follow the streamlines that take toward the origin, otherwise, a significant control effort will be presented to guarantee the convergence to the target.

%--disturbances at execution time--  

\subsection{Estimate of USV position and heading}

Learning just positions is not enough to reproduce the real displacement of the USV over the sea; therefore, the variable heading must be considered. Our second set of experiments is conducted to estimate not only position, but heading as well. 
%The USV trajectory is inherently tied to position and heading, hence our method must learn three variables instead of two (like the previous set of experiments). 
The USV trajectory is inherently tied to these three variables, hence our method must learn and guarantee stability over a 3D trajectory instead of a 2D trajectory (like the previous set of experiments).

\begin{figure}[htbp]
    \centering 
    \includegraphics[width=1\textwidth]{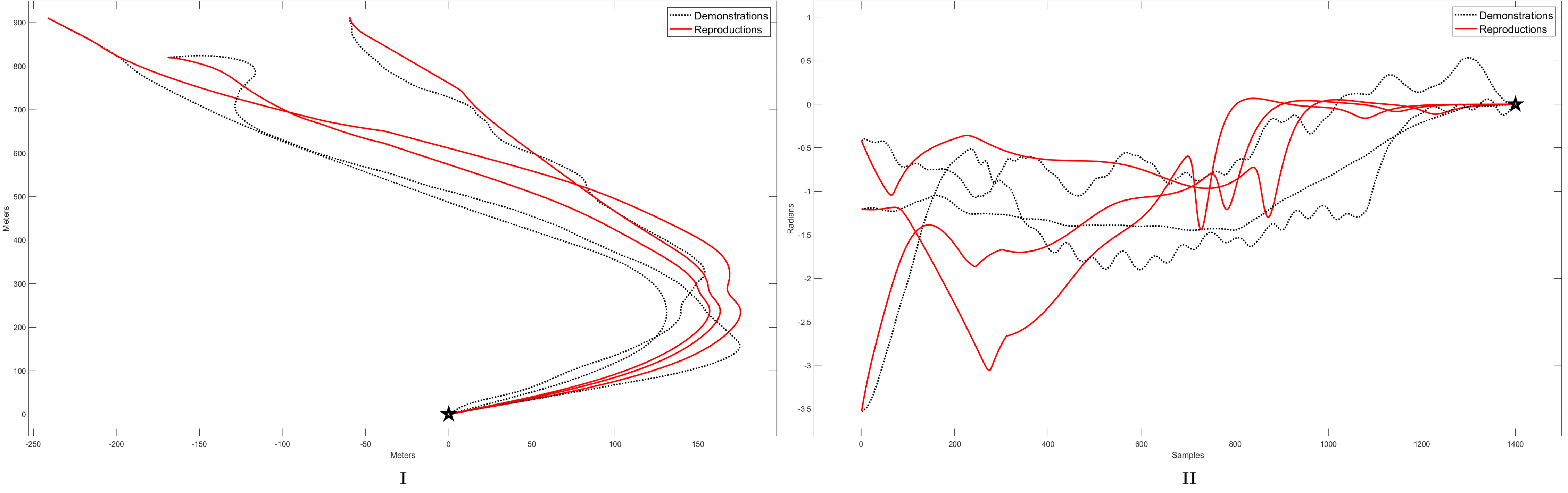}
    \caption{\small Estimate of USV position and heading. I) USV position in meters II) USV heading in radians}
    \label{fig:Estimate3D}
\end{figure}

The heading variable changes drastically in comparison with position variables, which makes a more complex dynamics to learn and stabilize. Therefore, a higher number of Gaussians is necessary to carry out a proper estimate. An estimate of the trajectory $A$ (position-heading) is depicted in Fig. \ref{fig:Estimate3D}, where $K=12$.  Note that the position estimate is qualitatively similar to the one achieved in the previous subsection, using $K=5$. 

The dynamics complexity, the number of variables to be learned, the number of data points as well as the number of demonstrations define the time for the learning process. Polar coordinates are used to reduce the data dimension and thus to simplify the computational complexity.  

\begin{figure}[htbp]
    \centering 
    \includegraphics[width=1\textwidth]{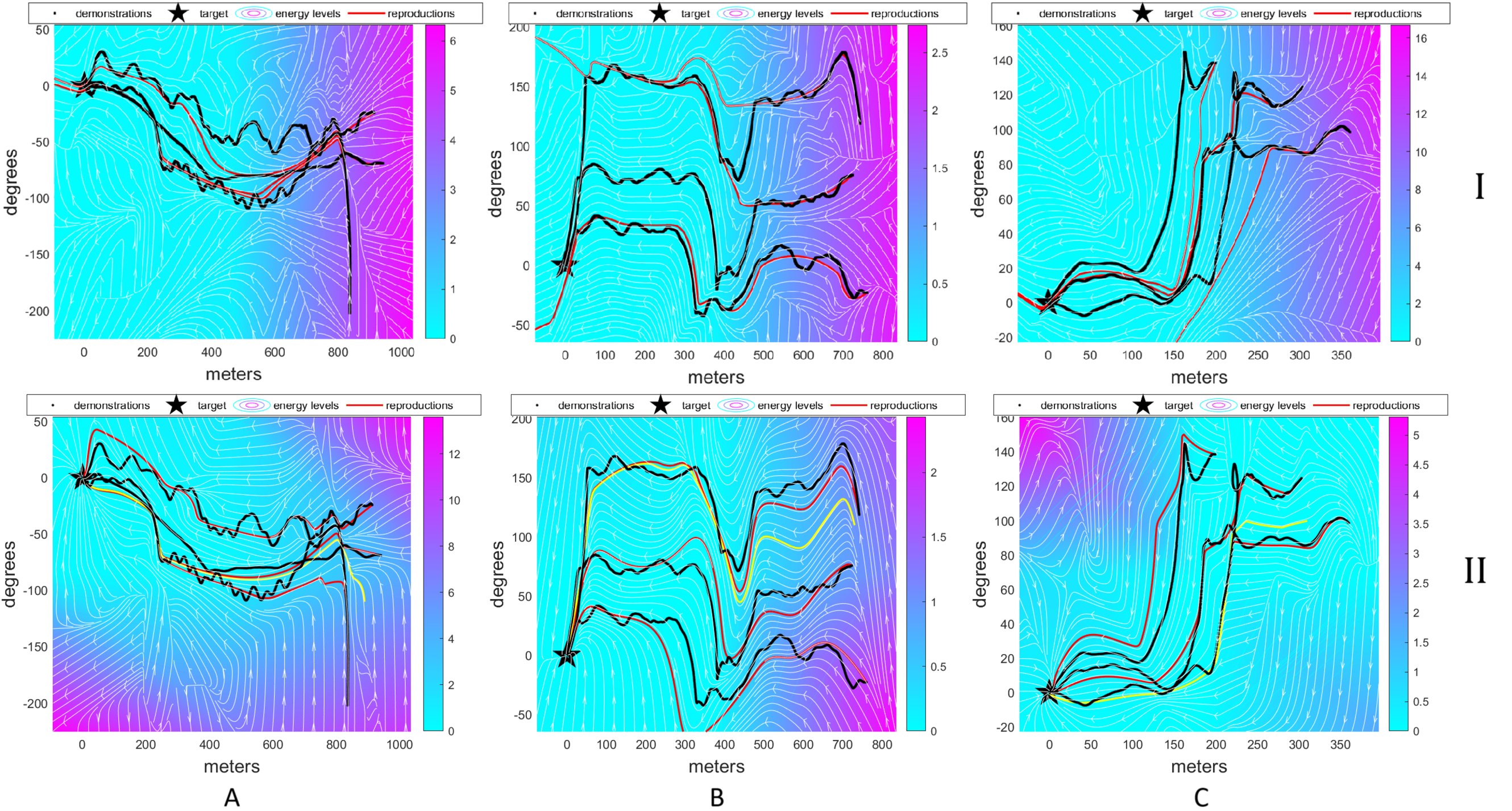}
    \caption{\small Estimates of USV position and heading in polar coordinates. I) GMR (No control) II) GMR + Control signal (Our method)}
    \label{fig:ZTheta_GMRControl}
\end{figure}

A 3D trajectory is converted to a 2D trajectory using the magnitude of the USV position and its heading. An estimate of the resulting demonstrations is calculated using just GMM/GMR without control and with $K=12$. As expected, such estimates are not good enough as all of them diverge from the target (see Fig. \ref{fig:ZTheta_GMRControl}-I). Note that the USV trajectories are more complex than the one presented above in Fig. \ref{fig:GMR_2D}, hence a greater number of Gaussian functions $K$ is necessary.

The proposed control law in our method improves the estimates of the trajectories and guarantee the convergence to the target using the same value of $K$ (see Fig. \ref{fig:ZTheta_GMRControl}-II). The solid white lines stand for streamlines and the background color depicts the energy-like function \eqref{eq:V1}: the darker the color, the higher the energy. The solid yellow line stand for a random initial condition to validate the effectiveness of our method when the UAV starts from a random position/heading. Note that the disturbances generate by the environment are rejected; nevertheless, relevant changes in dynamics remain.    

%\begin{figure}[htbp]
    %\centering 
    %\includegraphics[width=1\textwidth]{Trajectories_Polar_USV.eps}
    %\caption{\small Estimates of USV position and heading using our method}
    %\label{fig:ZThetaOurs}
%\end{figure}

\subsection{Error quantification}

\begin{figure}[htbp]
    \centering 
    \includegraphics[width=0.75\textwidth]{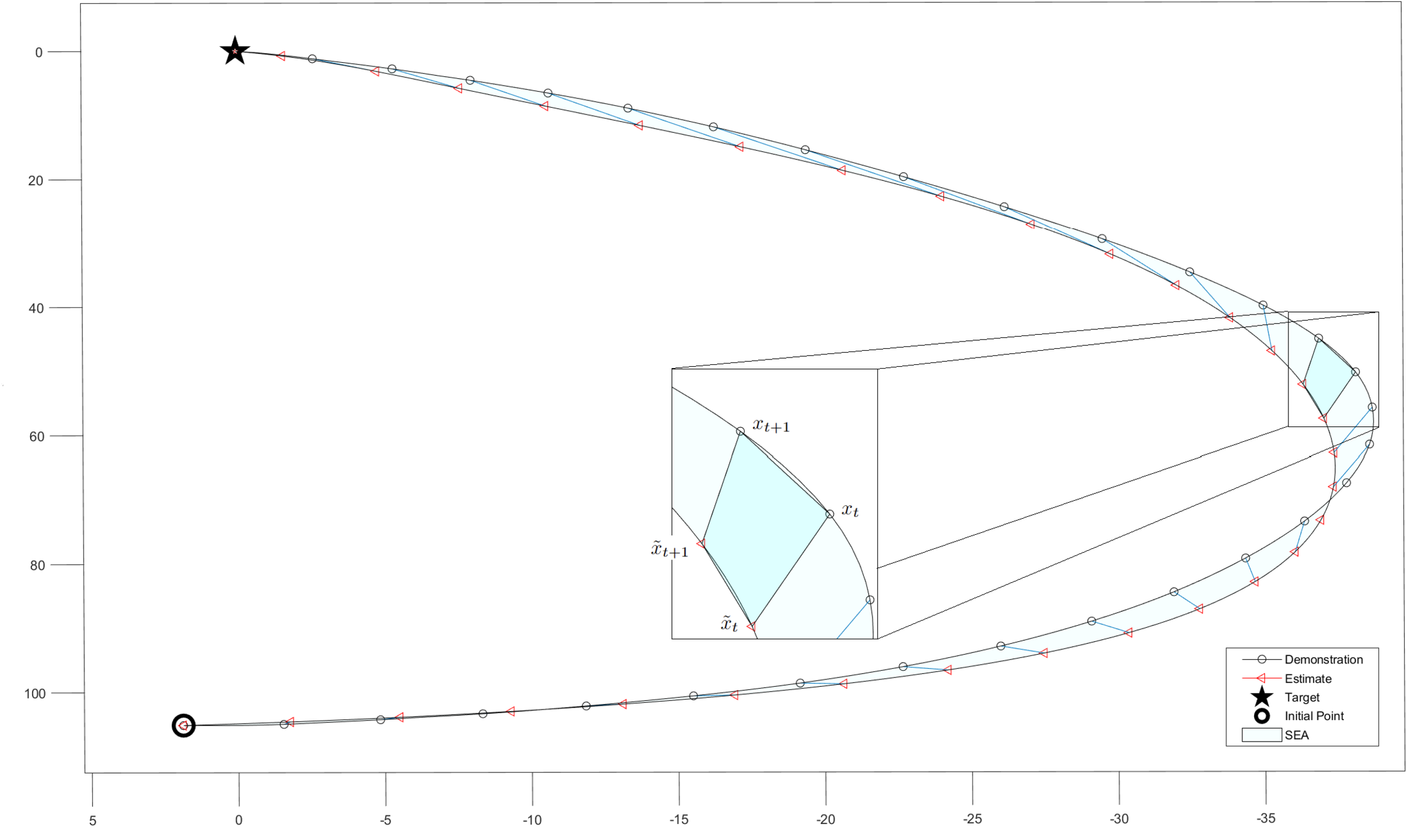}
    \caption{\small Illustrative example of the Swept Error Area (SEA)}
    \label{fig:SEA}
\end{figure}

The difference between the demonstrations and the estimates are measured through the Swept Error Area (SEA) \cite{r8}. This error can be understood as the area of an irregular shape generated by two curves with the same initial and final points; hence, the smaller the area, the better the estimate. This metric can be calculated by:

\begin{equation*}
%\label{eq:SEA}
    %\scalemath{0.75}
    SEA= {\sum_{t=0}^{T} A(x_t, x_{t+1}, \tilde{x}_t, \tilde{x}_{t+1}) },
\end{equation*}

where $A(\cdot)$ is the area of the tetragon defined by four sampling points $x_t$, $x_{t+1}$, $\tilde{x}_t$, $\tilde{x}_{t+1}$; which correspond to demonstration and estimate respectively (see Fig. \ref{fig:SEA}).  

\begin{figure}[htbp]
    \centering 
    \includegraphics[width=0.75\textwidth]{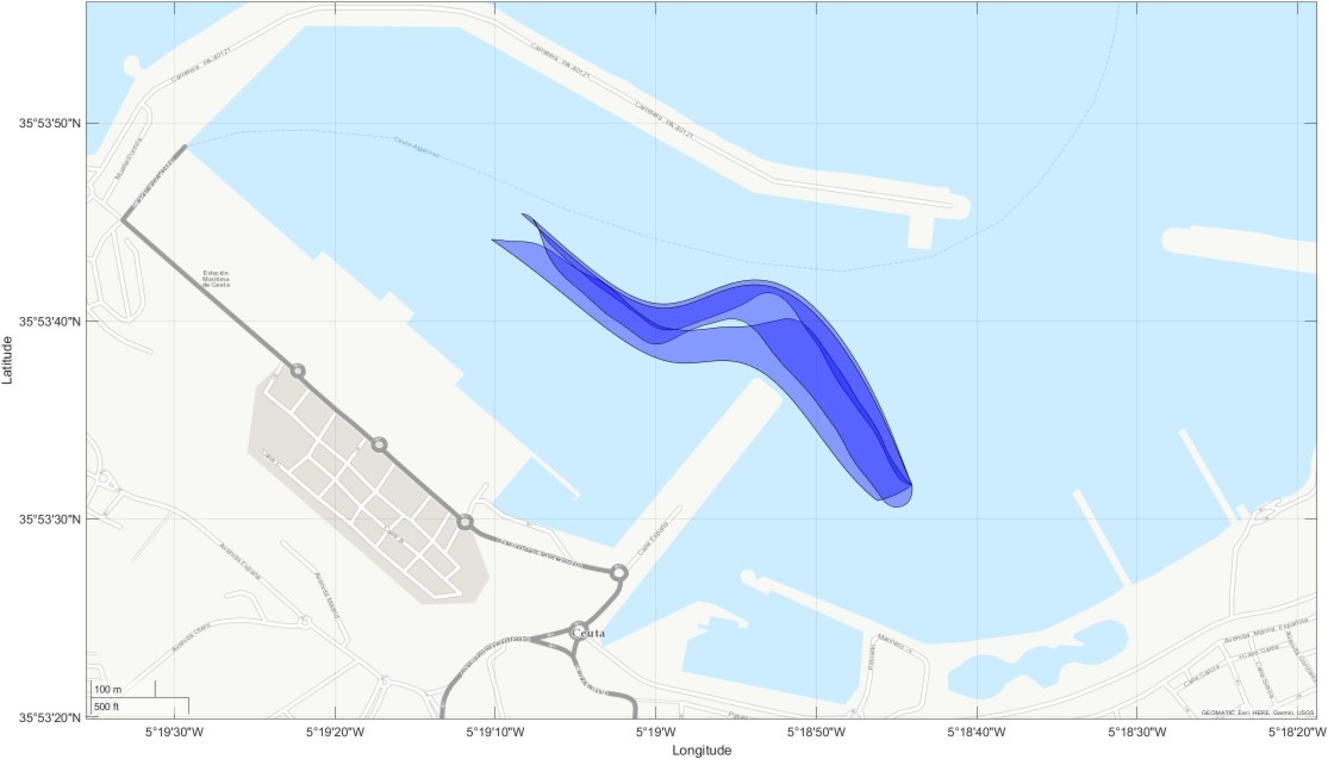}
    \caption{\small SEA of trajectory B}
    \label{fig:MarineAreaSEA}
\end{figure}

%The port of Ceuta has a marine area of approximately 9.76 $Km^2$
With a marine area of approximately 10 $Km^2$, the port of Ceuta served as an environment to validate the USV learning and stability under different trajectories and disturbances. The qualitative results show a great similarity between demonstrations and estimates (see Fig. \ref{fig:XY}), whereas the quantitative results, employing SEA are reported in Table \ref{tab:2}. An average error measures the difference among real and estimated trajectories, i.e. a marine area is calculated between both of them.
%
%where error measures for learning of position and heading/position are calculated for every trajectory.
%
\begin{table}
        \centering        
        \begin{tabular}{c|c|c|c|}
            \cline{2-4}
             & \multicolumn{3}{c|}{Trajectories}\\
            \cline{2-4}
             & A & B & C\\
            \hline
            \multicolumn{1}{|c|}{Area ($Km^2$)} & 0.0144 & 0.0374 &  0.0025\\
            \hline
        \end{tabular}
        \caption{SEA results}
        \label{tab:2}
    \end{table}
In order to illustrate the calculated SEA, Fig. \ref{fig:MarineAreaSEA} depicts the areas formed by the demonstrations and estimates (irregular, blue shapes) for the trajectory B. It is also worth noting that the total marine area formed by the three trajectories is actually a convergence zone where any initial state inside it, will follow the learned dynamics and guarantee the stability.

\subsection{Computational complexity}
The computational cost to learn the optimal parameters for $\bm \theta$ was mostly affected by the number of demonstrations $M$, the number of data points $N$, the data dimension $d$ and the number of Gaussians $K$. Furthermore, the considered operations to calculate it were scalar and matrix multiplication/division, matrix inversions and determinants. Thereby, the computational complexity of the proposed algorithm in this study is of order of magnitude $\mathcal{O}(8d^{21}KMN)$ per iteration in the optimization loop. The number of asymmetric quadratic function $L$ is irrelevant because adds only $\mathcal{O}(d^{6}L)$. The linear growth of the computational cost is intrinsically related to the size of $KMN$ while the exponential growth is related to the dimension of the trajectory (2D or 3D).
%
%, which is related to the dimension of the trajectory (2D or 3D).
%
\begin{table}
        \centering
        \begin{tabular}{c|c|c|c|}
            \cline{2-4}
             & \multicolumn{3}{c|}{Dimension}\\
            \cline{1-4}
            \multicolumn{1}{|c|}{Gaussians} & 2D  & 3D &  3D - polar\\
			%Gaussians & 2D & 3D & 3D - polar\\
            \hline
            \multicolumn{1}{|c|}{$K=5$} & 1  & N/A &  N/A\\
            \hline
            \multicolumn{1}{|c|}{$K=12$} & N/A & 46 &  10 \\
            \hline
        \end{tabular}
        \caption{Learning time in minutes}
        \label{tab:3}
    \end{table}
The learning time for the different trajectories is presented at Table \ref{tab:3}. The number of data points per every trajectory is between 1174 and 3411; the number of demonstrations is 3; the number of Gaussians is 5 or 12, depending on the dimension of the data; and finally the dimension is 2D or 3D. 

\section{Conclusion}

In this paper, we presented a novel data-driven learning and control approach for marine applications, to learn complex trajectories from human demonstrations and ensuring stability of multi-dimensional trajectories. The dynamics of the USV and the control law to converge to the target and/or reject disturbances are learned by solving a nonlinear constrained optimization problem.
Our approach was evaluated with real data coming from the Vendaval USV at the port of Ceuta; therefore, natural disturbances as wind and maritime current were considered to carry out the experiments. We presented a robust algorithm for USV that not only learns from noisy data, but recovers from big perturbations as well. Furthermore, our approach is able to learn USV position and heading at the same time.
The proper estimate of a trajectory depends on the right choice of parameters (the number of Gaussians $K$ and the number of asymmetric functions $L$). However, the more complex the trajectory, the greater the number of parameters is used.

Our future work will focus on finding new methods to estimate the dispersion of data and the loss of data-points. Besides, a more detailed study of a convergence zone will carry out instead of a single trajectory. We will also focus on integrating obstacle avoidance during the execution of the task. 

\section*{Acknowledgements}
The authors acknowledge support from the Projects 2024/00000362 for emerging 
research group Multi-Robot And Control Systems (MACS) under the framework VII PPIT-US 2024, and by the National Program for Doctoral Formation (Minciencias-Colombia, 885-2020).

\bibliography{references2}

\begin{thebibliography}{30}
\expandafter\ifx\csname natexlab\endcsname\relax\def\natexlab#1{#1}\fi
\providecommand{\url}[1]{\texttt{#1}}
\providecommand{\href}[2]{#2}
\providecommand{\path}[1]{#1}
\providecommand{\DOIprefix}{doi:}
\providecommand{\ArXivprefix}{arXiv:}
\providecommand{\URLprefix}{URL: }
\providecommand{\Pubmedprefix}{pmid:}
\providecommand{\doi}[1]{\href{http://dx.doi.org/#1}{\path{#1}}}
\providecommand{\Pubmed}[1]{\href{pmid:#1}{\path{#1}}}
\providecommand{\bibinfo}[2]{#2}
\ifx\xfnm\relax \def\xfnm[#1]{\unskip,\space#1}\fi
%Type = Article
\bibitem[{Acosta et~al.(2018)Acosta, D{\`o}ria-Cerezo and Fossas}]{Acosta}
\bibinfo{author}{Acosta, J.{\'A}.}, \bibinfo{author}{D{\`o}ria-Cerezo, A.}, \bibinfo{author}{Fossas, E.}, \bibinfo{year}{2018}.
\newblock \bibinfo{title}{Stabilisation of state-and-input constrained nonlinear systems via diffeomorphisms: A sontag's formula approach with an actual application}.
\newblock \bibinfo{journal}{International Journal of Robust and Nonlinear Control} \bibinfo{volume}{28}, \bibinfo{pages}{4032--4044}.
\newblock \DOIprefix\doi{https://doi.org/10.1002/rnc.4119}.
%Type = Article
\bibitem[{Artstein(1983)}]{Arstein}
\bibinfo{author}{Artstein, Z.}, \bibinfo{year}{1983}.
\newblock \bibinfo{title}{Stabilization with relaxed controls}.
\newblock \bibinfo{journal}{Nonlinear Analysis: Theory, Methods \& Applications} \bibinfo{volume}{7}, \bibinfo{pages}{1163--1173}.
\newblock \DOIprefix\doi{https://doi.org/10.1016/0362-546X(83)90049-4}.
%Type = Article
\bibitem[{Bai et~al.(2023)Bai, Li, Xu and Xiao}]{XIANGENBAI}
\bibinfo{author}{Bai, X.}, \bibinfo{author}{Li, B.}, \bibinfo{author}{Xu, X.}, \bibinfo{author}{Xiao, Y.}, \bibinfo{year}{2023}.
\newblock \bibinfo{title}{Usv path planning algorithm based on plant growth}.
\newblock \bibinfo{journal}{Ocean Engineering} \bibinfo{volume}{273}, \bibinfo{pages}{113965}.
\newblock \URLprefix \url{https://www.sciencedirect.com/science/article/pii/S0029801823003499}, \DOIprefix\doi{https://doi.org/10.1016/j.oceaneng.2023.113965}.
%Type = Article
\bibitem[{Becerra-Mora and Acosta(2024)}]{BecAco24}
\bibinfo{author}{Becerra-Mora, Y.A.}, \bibinfo{author}{Acosta, J.{\'A}.}, \bibinfo{year}{2024}.
\newblock \bibinfo{title}{Data-driven learning and control of nonlinear system dynamics}.
\newblock \bibinfo{journal}{Nonlinear Dynamics} \URLprefix \url{https://doi.org/10.1007/s11071-024-10149-1}, \DOIprefix\doi{10.1007/s11071-024-10149-1}.
%Type = Article
\bibitem[{Birk et~al.(2018)Birk, Doernbach, Mueller, {\L}uczynski, Gomez~Chavez, Koehntopp, Kupcsik, Calinon, Tanwani, Antonelli, Di~Lillo, Simetti, Casalino, Indiveri, Ostuni, Turetta, Caffaz, Weiss, Gobert, Chemisky, Gancet, Siedel, Govindaraj, Martinez and Letier}]{BIRK}
\bibinfo{author}{Birk, A.}, \bibinfo{author}{Doernbach, T.}, \bibinfo{author}{Mueller, C.}, \bibinfo{author}{{\L}uczynski, T.}, \bibinfo{author}{Gomez~Chavez, A.}, \bibinfo{author}{Koehntopp, D.}, \bibinfo{author}{Kupcsik, A.}, \bibinfo{author}{Calinon, S.}, \bibinfo{author}{Tanwani, A.K.}, \bibinfo{author}{Antonelli, G.}, \bibinfo{author}{Di~Lillo, P.}, \bibinfo{author}{Simetti, E.}, \bibinfo{author}{Casalino, G.}, \bibinfo{author}{Indiveri, G.}, \bibinfo{author}{Ostuni, L.}, \bibinfo{author}{Turetta, A.}, \bibinfo{author}{Caffaz, A.}, \bibinfo{author}{Weiss, P.}, \bibinfo{author}{Gobert, T.}, \bibinfo{author}{Chemisky, B.}, \bibinfo{author}{Gancet, J.}, \bibinfo{author}{Siedel, T.}, \bibinfo{author}{Govindaraj, S.}, \bibinfo{author}{Martinez, X.}, \bibinfo{author}{Letier, P.}, \bibinfo{year}{2018}.
\newblock \bibinfo{title}{Dexterous underwater manipulation from onshore locations: Streamlining efficiencies for remotely operated underwater vehicles}.
\newblock \bibinfo{journal}{IEEE Robotics \& Automation Magazine} \bibinfo{volume}{25}, \bibinfo{pages}{24--33}.
\newblock \DOIprefix\doi{10.1109/MRA.2018.2869523}.
%Type = Book
\bibitem[{Bishop(2006)}]{Bishop}
\bibinfo{author}{Bishop, C.}, \bibinfo{year}{2006}.
\newblock \bibinfo{title}{Patter Recognition and Machine Learning}.
\newblock \bibinfo{publisher}{Springer New York, NY}.
%Type = Book
\bibitem[{Calinon(2009)}]{Calinon09book}
\bibinfo{author}{Calinon, S.}, \bibinfo{year}{2009}.
\newblock \bibinfo{title}{Robot Programming by Demonstration: A Probabilistic Approach}.
\newblock \bibinfo{publisher}{EPFL/CRC Press}.
\newblock \bibinfo{note}{EPFL Press ISBN 978-2-940222-31-5, CRC Press ISBN 978-1-4398-0867-2}.
%Type = Article
\bibitem[{Chaysri et~al.(2023)Chaysri, Spatharis, Blekas and Vlachos}]{CHAYSRI}
\bibinfo{author}{Chaysri, P.}, \bibinfo{author}{Spatharis, C.}, \bibinfo{author}{Blekas, K.}, \bibinfo{author}{Vlachos, K.}, \bibinfo{year}{2023}.
\newblock \bibinfo{title}{Unmanned surface vehicle navigation through generative adversarial imitation learning}.
\newblock \bibinfo{journal}{Ocean Engineering} \bibinfo{volume}{282}, \bibinfo{pages}{114989}.
\newblock \URLprefix \url{https://www.sciencedirect.com/science/article/pii/S0029801823013732}, \DOIprefix\doi{https://doi.org/10.1016/j.oceaneng.2023.114989}.
%Type = Article
\bibitem[{Cohn et~al.(1996)Cohn, Ghahramani and Jordan}]{r23}
\bibinfo{author}{Cohn, D.A.}, \bibinfo{author}{Ghahramani, Z.}, \bibinfo{author}{Jordan, M.I.}, \bibinfo{year}{1996}.
\newblock \bibinfo{title}{Active learning with statistical models}.
\newblock \bibinfo{journal}{Journal of Artificial Intelligence Research} \bibinfo{volume}{4}, \bibinfo{pages}{129--145}.
%Type = Article
\bibitem[{Dempster et~al.(1977)Dempster, Laird and Rubin}]{r22}
\bibinfo{author}{Dempster, A.P.}, \bibinfo{author}{Laird, N.M.}, \bibinfo{author}{Rubin, D.B.}, \bibinfo{year}{1977}.
\newblock \bibinfo{title}{Maximum likelihood from incomplete data via the em algorithm}.
\newblock \bibinfo{journal}{Journal of the Royal Statistical Society} \bibinfo{volume}{39}, \bibinfo{pages}{1--38}.
%Type = Article
\bibitem[{Deraj et~al.(2023)Deraj, Kumar, Alam and Somayajula}]{DERAJ}
\bibinfo{author}{Deraj, R.}, \bibinfo{author}{Kumar, R.S.}, \bibinfo{author}{Alam, M.S.}, \bibinfo{author}{Somayajula, A.}, \bibinfo{year}{2023}.
\newblock \bibinfo{title}{Deep reinforcement learning based controller for ship navigation}.
\newblock \bibinfo{journal}{Ocean Engineering} \bibinfo{volume}{273}, \bibinfo{pages}{113937}.
\newblock \URLprefix \url{https://www.sciencedirect.com/science/article/pii/S0029801823003219}, \DOIprefix\doi{https://doi.org/10.1016/j.oceaneng.2023.113937}.
%Type = Inproceedings
\bibitem[{Farchy et~al.(2013)Farchy, Barrett, MacAlpine and Stone}]{FARCHY}
\bibinfo{author}{Farchy, A.}, \bibinfo{author}{Barrett, S.}, \bibinfo{author}{MacAlpine, P.}, \bibinfo{author}{Stone, P.}, \bibinfo{year}{2013}.
\newblock \bibinfo{title}{Humanoid robots learning to walk faster: From the real world to simulation and back}, in: \bibinfo{booktitle}{Proceedings of the 2013 International Conference on Autonomous Agents and Multi-Agent Systems}, \bibinfo{publisher}{International Foundation for Autonomous Agents and Multiagent Systems}, \bibinfo{address}{Richland, SC}. pp. \bibinfo{pages}{39--46}.
%Type = Article
\bibitem[{Gonzalez-Garcia et~al.(2022)Gonzalez-Garcia, Collado-Gonzalez, Cuan-Urquizo, Sotelo, Sotelo and Casta{\~n}eda}]{GONZALEZGARCIA}
\bibinfo{author}{Gonzalez-Garcia, A.}, \bibinfo{author}{Collado-Gonzalez, I.}, \bibinfo{author}{Cuan-Urquizo, R.}, \bibinfo{author}{Sotelo, C.}, \bibinfo{author}{Sotelo, D.}, \bibinfo{author}{Casta{\~n}eda, H.}, \bibinfo{year}{2022}.
\newblock \bibinfo{title}{Path-following and lidar-based obstacle avoidance via nmpc for an autonomous surface vehicle}.
\newblock \bibinfo{journal}{Ocean Engineering} \bibinfo{volume}{266}, \bibinfo{pages}{112900}.
\newblock \URLprefix \url{https://www.sciencedirect.com/science/article/pii/S0029801822021837}, \DOIprefix\doi{https://doi.org/10.1016/j.oceaneng.2022.112900}.
%Type = Misc
\bibitem[{IALA(2024)}]{IALA}
\bibinfo{author}{IALA}, \bibinfo{year}{2024}.
\newblock \bibinfo{title}{International organization for marine aids to navigation}.
\newblock \URLprefix \url{https://www.iala.int/}. \bibinfo{note}{last accessed 27 July 2024}.
%Type = Article
\bibitem[{Khansari and Billard(2014)}]{r8}
\bibinfo{author}{Khansari, S.M.}, \bibinfo{author}{Billard, A.}, \bibinfo{year}{2014}.
\newblock \bibinfo{title}{Learning control lyapunov function to ensure stability of dynamical system-based robot reaching motions}.
\newblock \bibinfo{journal}{Robotics and Autonomous Systems} \bibinfo{volume}{62}, \bibinfo{pages}{752--765}.
%Type = Inproceedings
\bibitem[{Li et~al.(2017)Li, Song and Ermon}]{LI2017}
\bibinfo{author}{Li, Y.}, \bibinfo{author}{Song, J.}, \bibinfo{author}{Ermon, S.}, \bibinfo{year}{2017}.
\newblock \bibinfo{title}{Infogail: Interpretable imitation learning from visual demonstrations}, in: \bibinfo{booktitle}{Proceedings of the 31st International Conference on Neural Information Processing Systems}, \bibinfo{publisher}{Curran Associates Inc.}, \bibinfo{address}{Red Hook, NY, USA}. pp. \bibinfo{pages}{3815--3825}.
%Type = Article
\bibitem[{Loquercio et~al.(2018)Loquercio, Maqueda, del Blanco and Scaramuzza}]{LOQUERCIO}
\bibinfo{author}{Loquercio, A.}, \bibinfo{author}{Maqueda, A.I.}, \bibinfo{author}{del Blanco, C.R.}, \bibinfo{author}{Scaramuzza, D.}, \bibinfo{year}{2018}.
\newblock \bibinfo{title}{Dronet: Learning to fly by driving}.
\newblock \bibinfo{journal}{IEEE Robotics and Automation Letters} \bibinfo{volume}{3}, \bibinfo{pages}{1088--1095}.
\newblock \DOIprefix\doi{10.1109/LRA.2018.2795643}.
%Type = Article
\bibitem[{Meng et~al.(2023)Meng, Humne, Bucknall, Englot and Liu}]{MENG}
\bibinfo{author}{Meng, J.}, \bibinfo{author}{Humne, A.}, \bibinfo{author}{Bucknall, R.}, \bibinfo{author}{Englot, B.}, \bibinfo{author}{Liu, Y.}, \bibinfo{year}{2023}.
\newblock \bibinfo{title}{A fully-autonomous framework of unmanned surface vehicles in maritime environments using gaussian process motion planning}.
\newblock \bibinfo{journal}{IEEE Journal of Oceanic Engineering} \bibinfo{volume}{48}, \bibinfo{pages}{59--79}.
\newblock \DOIprefix\doi{10.1109/JOE.2022.3194165}.
%Type = Article
\bibitem[{Ouyang et~al.(2023)Ouyang, Chen and Zou}]{OUYANG}
\bibinfo{author}{Ouyang, Z.L.}, \bibinfo{author}{Chen, G.}, \bibinfo{author}{Zou, Z.J.}, \bibinfo{year}{2023}.
\newblock \bibinfo{title}{Identification modeling of ship maneuvering motion based on local gaussian process regression}.
\newblock \bibinfo{journal}{Ocean Engineering} \bibinfo{volume}{267}, \bibinfo{pages}{113251}.
\newblock \URLprefix \url{https://www.sciencedirect.com/science/article/pii/S0029801822025343}, \DOIprefix\doi{https://doi.org/10.1016/j.oceaneng.2022.113251}.
%Type = Article
\bibitem[{Shah and Gupta(2020)}]{BRUAL}
\bibinfo{author}{Shah, B.C.}, \bibinfo{author}{Gupta, S.K.}, \bibinfo{year}{2020}.
\newblock \bibinfo{title}{Long-distance path planning for unmanned surface vehicles in complex marine environment}.
\newblock \bibinfo{journal}{IEEE Journal of Oceanic Engineering} \bibinfo{volume}{45}, \bibinfo{pages}{813--830}.
\newblock \DOIprefix\doi{10.1109/JOE.2019.2909508}.
%Type = Article
\bibitem[{Sontag(1989)}]{Sontag}
\bibinfo{author}{Sontag, E.}, \bibinfo{year}{1989}.
\newblock \bibinfo{title}{A `universal' construction of artstein's theorem on nonlinear stabilization}.
\newblock \bibinfo{journal}{Systems \& Control Letters} \bibinfo{volume}{13}, \bibinfo{pages}{117--123}.
\newblock \DOIprefix\doi{https://doi.org/10.1016/0167-6911(89)90028-5}.
%Type = Inproceedings
\bibitem[{Vogt et~al.(2017)Vogt, Stepputtis, Grehl, Jung and Ben~Amor}]{VOGT}
\bibinfo{author}{Vogt, D.}, \bibinfo{author}{Stepputtis, S.}, \bibinfo{author}{Grehl, S.}, \bibinfo{author}{Jung, B.}, \bibinfo{author}{Ben~Amor, H.}, \bibinfo{year}{2017}.
\newblock \bibinfo{title}{A system for learning continuous human-robot interactions from human-human demonstrations}, in: \bibinfo{booktitle}{2017 IEEE International Conference on Robotics and Automation (ICRA)}, pp. \bibinfo{pages}{2882--2889}.
\newblock \DOIprefix\doi{10.1109/ICRA.2017.7989334}.
%Type = Article
\bibitem[{Wang et~al.(2019)Wang, Jin and Er}]{NINGWANG}
\bibinfo{author}{Wang, N.}, \bibinfo{author}{Jin, X.}, \bibinfo{author}{Er, M.J.}, \bibinfo{year}{2019}.
\newblock \bibinfo{title}{A multilayer path planner for a usv under complex marine environments}.
\newblock \bibinfo{journal}{Ocean Engineering} \bibinfo{volume}{184}, \bibinfo{pages}{1--10}.
\newblock \URLprefix \url{https://www.sciencedirect.com/science/article/pii/S0029801819302422}, \DOIprefix\doi{https://doi.org/10.1016/j.oceaneng.2019.05.017}.
%Type = Article
\bibitem[{Wang and Xu(2020)}]{Wang20}
\bibinfo{author}{Wang, N.}, \bibinfo{author}{Xu, H.}, \bibinfo{year}{2020}.
\newblock \bibinfo{title}{Dynamics-constrained global-local hybrid path planning of an autonomous surface vehicle}.
\newblock \bibinfo{journal}{IEEE Transactions on Vehicular Technology} \bibinfo{volume}{69}, \bibinfo{pages}{6928--6942}.
\newblock \DOIprefix\doi{10.1109/TVT.2020.2991220}.
%Type = Article
\bibitem[{Wang et~al.(2023)Wang, Liu, Tian, Zhang, Qiao and Wang}]{PENGWANG}
\bibinfo{author}{Wang, P.}, \bibinfo{author}{Liu, R.}, \bibinfo{author}{Tian, X.}, \bibinfo{author}{Zhang, X.}, \bibinfo{author}{Qiao, L.}, \bibinfo{author}{Wang, Y.}, \bibinfo{year}{2023}.
\newblock \bibinfo{title}{Obstacle avoidance for environmentally-driven usvs based on deep reinforcement learning in large-scale uncertain environments}.
\newblock \bibinfo{journal}{Ocean Engineering} \bibinfo{volume}{270}, \bibinfo{pages}{113670}.
\newblock \URLprefix \url{https://www.sciencedirect.com/science/article/pii/S0029801823000549}, \DOIprefix\doi{https://doi.org/10.1016/j.oceaneng.2023.113670}.
%Type = Article
\bibitem[{Wei et~al.(2022)Wei, Bai, Wei, Ji and Liu}]{WEI}
\bibinfo{author}{Wei, C.}, \bibinfo{author}{Bai, H.}, \bibinfo{author}{Wei, Y.}, \bibinfo{author}{Ji, Z.}, \bibinfo{author}{Liu, Z.}, \bibinfo{year}{2022}.
\newblock \bibinfo{title}{Learning manipulation skills with demonstrations for the swing process control of dredgers}.
\newblock \bibinfo{journal}{Ocean Engineering} \bibinfo{volume}{246}, \bibinfo{pages}{110545}.
\newblock \URLprefix \url{https://www.sciencedirect.com/science/article/pii/S0029801822000233}, \DOIprefix\doi{https://doi.org/10.1016/j.oceaneng.2022.110545}.
%Type = Article
\bibitem[{Wright(2005)}]{Wright}
\bibinfo{author}{Wright, M.H.}, \bibinfo{year}{2005}.
\newblock \bibinfo{title}{The interior-point revolution in optimization: {H}istory, recent developments, and lasting consequences}.
\newblock \bibinfo{journal}{Bull. Amer. Math. Soc.} \bibinfo{volume}{42}, \bibinfo{pages}{39--56}.
%Type = Article
\bibitem[{Xu et~al.(2023)Xu, Qin, Ma, Deng and Xue}]{PEILONG}
\bibinfo{author}{Xu, P.}, \bibinfo{author}{Qin, H.}, \bibinfo{author}{Ma, J.}, \bibinfo{author}{Deng, Z.}, \bibinfo{author}{Xue, Y.}, \bibinfo{year}{2023}.
\newblock \bibinfo{title}{Data-driven model predictive control for ships with gaussian process}.
\newblock \bibinfo{journal}{Ocean Engineering} \bibinfo{volume}{268}, \bibinfo{pages}{113420}.
\newblock \URLprefix \url{https://www.sciencedirect.com/science/article/pii/S0029801822027032}, \DOIprefix\doi{https://doi.org/10.1016/j.oceaneng.2022.113420}.
%Type = Article
\bibitem[{Xue et~al.(2020)Xue, Liu, Ji, Xue and Huang}]{YIFANXUE}
\bibinfo{author}{Xue, Y.}, \bibinfo{author}{Liu, Y.}, \bibinfo{author}{Ji, C.}, \bibinfo{author}{Xue, G.}, \bibinfo{author}{Huang, S.}, \bibinfo{year}{2020}.
\newblock \bibinfo{title}{System identification of ship dynamic model based on gaussian process regression with input noise}.
\newblock \bibinfo{journal}{Ocean Engineering} \bibinfo{volume}{216}, \bibinfo{pages}{107862}.
\newblock \URLprefix \url{https://www.sciencedirect.com/science/article/pii/S0029801820308301}, \DOIprefix\doi{https://doi.org/10.1016/j.oceaneng.2020.107862}.
%Type = Article
\bibitem[{Zhang et~al.(2023)Zhang, Shang, Liu and Zhang}]{ZHANG2023}
\bibinfo{author}{Zhang, G.}, \bibinfo{author}{Shang, X.}, \bibinfo{author}{Liu, J.}, \bibinfo{author}{Zhang, W.}, \bibinfo{year}{2023}.
\newblock \bibinfo{title}{Improved iterative learning path-following control for usv via the potential-based dvs guidance}.
\newblock \bibinfo{journal}{Ocean Engineering} \bibinfo{volume}{280}, \bibinfo{pages}{114543}.
\newblock \URLprefix \url{https://www.sciencedirect.com/science/article/pii/S0029801823009277}, \DOIprefix\doi{https://doi.org/10.1016/j.oceaneng.2023.114543}.

\end{thebibliography}
\bibliographystyle{elsarticle-harv}

\end{document}